\documentclass{article}

\usepackage{graphicx}
\usepackage{float}
\usepackage{citesort}
\usepackage{amsmath,amssymb}
\usepackage{color}
\usepackage{caption}
\usepackage{wrapfig}
\usepackage{wasysym}

\begin{document}

\title{Excitation function of elastic $pp$ scattering\\from a unitarily extended Bialas-Bzdak model}

\author{F. Nemes$^{1,2}$ \\ $^1$ CERN, CH-1211 Geneva 23, Switzerland\\
frigyes.janos.nemes@cern.ch\vspace{5mm}\\ 
T. Cs\"org\H{o}$^{2,3}$\\ $^2$ Wigner Research Centre for Physics, Hungarian Academy of Sciences\\ H-1525 Budapest 114, P.O.Box 49, Hungary\\
$^3$K\'aroly R\'obert College, H-3200 Gy\"ongy\"os, M\'atrai \'ut 36, Hungary\\
csorgo.tamas@wigner.mta.hu\vspace{5mm}\\
M. Csan\'ad$^{4}$\\ $^{4}$ E\"otv\"os University, Department of Atomic Physics\\ H-1117 Budapest, P\'azm\'any P\'eter s., 1/A Hungary\\
csanad@elte.hu\vspace{5mm}}

\maketitle

\begin{abstract} 
The Bialas-Bzdak model of elastic proton-proton scattering
assumes a purely imaginary forward scattering amplitude, which consequently
vanishes at the diffractive minima.  We extended the model to arbitrarily large
real parts in a way that constraints from unitarity are satisfied.  The
resulting model is able to describe elastic $pp$ scattering not only at the
lower ISR energies but also at $\sqrt{s}=$7~TeV in a statistically acceptable
manner, both in the diffractive cone and in the region of the first diffractive
minimum.  The total cross-section as well as the differential cross-section of
elastic proton-proton scattering is predicted for the future LHC energies of
$\sqrt{s}=$13, 14, 15~TeV and also to 28~TeV.  A non-trivial, significantly
non-exponential feature of the differential cross-section of elastic
proton-proton scattering is analyzed and the excitation function of the
non-exponential behavior is predicted. The excitation function of the shadow
profiles is discussed and related to saturation at small impact parameters. 

\end{abstract}

\section{Introduction}

In a pair of recent papers the Bialas-Bzdak model (BB)~\cite{Bialas:2006qf} of small angle elastic proton-proton ($pp$) scattering at high energies was studied at $\sqrt{s}=7$~TeV center-of-mass LHC energy~\cite{CsorgO:2013kua,Nemes:2012cp}. 

In this work, we extend those investigations by improving on 
the original BB model by adding a real part to its forward scattering amplitude (FSA) in a unitary manner, and, furthermore, we present the extrapolation of the BB model  to future LHC energies as well.  Our method to include the energy evolution of
the parameters is similar to the so-called ``geometric scaling'' discussed in Refs.~\cite{Ferreira:2014gda} and~\cite{Kohara:2014waa},
however, in our case only the constant and the linear terms are used as a function of $\ln\left(s\right)$, while the quadratic terms used in
Refs.~\cite{Ferreira:2014gda} and \cite{Kohara:2014waa} were not needed at our current level of precision.

During 2014, 
the TOTEM experiment made  public an important preliminary experimental
observation at $\sqrt{s}=8$~TeV LHC energy: the $pp$ elastic differential cross-section
shows a deviation from the simplest, exponential behavior at low-$t$, where $t$ is the
squared four-momentum transfer of the $pp$ scattering process~\cite{Simone:WPCF2014}. This feature of
the $\sqrt{s}=8$~TeV preliminary TOTEM dataset was related to \mbox{$t$-channel}
unitarity of the FSA in
Ref.~\cite{Jenkovszky:2014yea}, a concept that we also focus on, using and
generalizing the quark-diquark model of Bialas and Bzdak
to determine the shape of the FSA in elastic $pp$ scattering.

In its original form, the BB model~\cite{Bialas:2006qf} assumes that the
real part of the FSA is negligible, correspondingly, the FSA vanishes at the diffractive minima. At the ISR energies of $\sqrt{s}=$23.5$-$62.5 GeV, that were first analyzed in the inspiring paper of  Bialas and Bzdak~\cite{Bialas:2006qf}, this assumption is indeed reasonable, as it is confirmed in Ref.~\cite{Nemes:2012cp}. At these ISR energies, only very few data points were available in the dip region around the first diffractive minimum of elastic $pp$ scattering, which were then left out from the BB model fits of Ref.~\cite{Nemes:2012cp} to achieve a quality description of the remaining data points. However, in recent years, TOTEM data~\cite{Antchev:2011vs} explored the dip region at the LHC energy of $\sqrt{s}=7$~TeV in
great details, at several different values of the squared four-momentum transfer $t$. Ref.~\cite{Nemes:2012cp} demonstrated, that the original BB model cannot
describe this dip region, not without at least a small real part that has to be
added to its FSA in a reasonable way.

Subsequently, the BB model has been generalized in Ref.~\cite{CsorgO:2013kua} 
by allowing for a perturbatively small real part of the FSA, which improved the agreement of the model with TOTEM data on elastic $pp$ scattering at the LHC energy of $\sqrt{s} = 7$~TeV.  It was expected that the main reason for the appearance of this real part is that certain rare elastic scattering of the constituents of the protons may be non-collinear thus may lead to inelastic events even if the elementary interactions are elastic.   The corresponding phenomenological generalization of the Bialas-Bzdak model~\cite{CsorgO:2013kua} was based on the assumption that the real part of the FSA is small, and can be handled perturbatively.
The resulting $\alpha$-generalized Bialas-Bzdak ($\alpha$BB) model was compared to ISR data in Ref.~\cite{CsorgO:2013kua}, and it was demonstrated that a small,  of the order of 1~\permil~real part of the FSA indeed  results in excellent
fit qualities and a statistically acceptable description of the data in the region of the diffractive minimum or dip. However, at the LHC energy of $\sqrt{s}=7$~TeV, although the real part of the fit becomes significantly larger than at ISR, 
the same $\alpha$BB model does not result in a satisfactory, statistically acceptable fit quality, although the visual quality of the fitted curves improve 
significantly as compared to that of the original BB model~\cite{CsorgO:2013kua}.
	
These results indicate that at the LHC energies the real part of the
FSA may reach significant values where unitarity constraints may already play an important role. The unitarity of the $S$-matrix provides also the basis for the optical theorem, which in turn provides a method to determine the total cross-section from  an extrapolation of the elastic scattering measurements to the $t = 0$ point. In the $\alpha$BB model~\cite{CsorgO:2013kua}, 
unitarity constraints were not explicitly considered: as the original BB model
with zero real part obeyed unitarity, adding a small real part may result only in small (and expected) deviations from unitarity and the optical theorem.  However, when the model was fitted to the 7~TeV TOTEM data in the dip region~\cite{CsorgO:2013kua}, the extrapolation to the point of $t = 0$ 
and the related value of the total cross-section underestimated the measured
total cross-section by about 40\%, suggesting, that perhaps the real part of
the FSA may be large, and unitarity relations should be explicitly considered. 

These indications motivate the present manuscript, where the BB model is further generalized to arbitrarily large real parts of the FSA, derived from unitarity constraints.  The resulting model is referred to as the real extended Bialas-Bzdak (ReBB) model.

In Refs.~\cite{Nemes:2012cp} and \cite{CsorgO:2013kua} a terse
overview is reported about the field of elastic scattering at high energies,
that summarizes the developments up to 2013. 
In this introduction let us highlight only some
more recent works, in order to put our results to a broader context
of recent, independent investigations of elastic $pp$ collisions
at the LHC energies.

The recent analysis of Ref.~\cite{Grichine:2012ry} applies a quark-diquark representation of the proton to describe TOTEM data~\cite{Antchev:2011vs}
also using antiproton-proton data. The quark-diquark approach of Ref.~\cite{Grichine:2012ry} is a continuation of studies from the late
sixties~\cite{Lichtenberg:1967zz}, which was first applied to describe $pp$ elastic scattering in Ref.~\cite{Tsarev:1978uc}. 
The early studies~\cite{Lichtenberg:1967zz,Tsarev:1978uc} provide the foundation of the quark-diquark representation of the BB model
as well. A recent improvement on this idea is Ref.~\cite{Grichine:2014wea}, which introduces the idea of ``Pomeron elasticity'' to increase the real part of
the FSA, which approach shows an interesting relationship with our present analysis.

The structure of this manuscript is as follows: in Section~\ref{sec:unitarity}, 
the general form of the FSA is re-derived for the case of
a non-vanishing real part starting from $S$-matrix unitarity. Then this result is applied to the extension of the BB model to a non-vanishing and possibly large
real part of the FSA. The mathematical relations between the resulting ReBB and the earlier $\alpha$BB models are specified in subsection~\ref{s:uBB}.

	In Section~\ref{sec:fitdescription}, the specified ReBB model is fitted to
TOTEM data on elastic $pp$ scattering at $\sqrt{s} = $ 7~TeV, both in the
diffractive cone~\cite{Antchev:2011zz,Antchev:2013gaa,Antchev:2013iaa} and in the dip region~\cite{Antchev:2011vs}.

Based on these fits and comparisons of the ReBB model to $\sqrt{s}=$ 7~TeV data, 
the shadow profile function $A(b)$ is evaluated
in Section~\ref{sec:shadow_profile}. This shadow profile function characterizes
the probability of inelastic $pp$ scattering at a given impact parameter~$b$, and is compared to the shadow profile functions of elastic $pp$ collisions at lower, ISR energies. Section \ref{sec:non_exponential} is devoted to study the structure of the differential cross-section $d\sigma/dt$ at low-$|t|$ values and also to compare it with a purely exponential behavior.

	In Section~\ref{sec:excitation}, the excitation function of 
the fit parameters is investigated and their evolution with $\sqrt{s}$ 
is obtained based on a geometrical picture. The model parameters are extrapolated 
to the expected future LHC energies of $\sqrt{s}=$13, 14 and 15 TeV, 
as well as for 28 TeV, that is not foreseen to be available 
at man-made accelerators in the  near future, but may be relevant 
for the investigation of cosmic ray events. The excitation functions of the shadow profile functions~$A(b)$
are also discussed. Finally we summarize and conclude.

\section{The real extended Bialas-Bzdak model}
\label{sec:unitarity} 

Although the original form of the Bialas-Bzdak model neglects the real part of
the FSA in high energy elastic $pp$ scattering, the
model is based on Glauber scattering theory and obeys unitarity constraints. 

The phenomenological generalization of the Bialas-Bzdak 
model~\cite{CsorgO:2013kua} is based on the assumption, that the
real part of the FSA is small, and can be handled
perturbatively.
However, it turned out that the addition of a small real part does not lead to a
statistically acceptable description of TOTEM data on elastic $pp$ collisions at~$\sqrt{s} = $ 7 TeV. 
In this manuscript, we consider the case, when the real part of the
FSA is not perturbatively small. 
We restart from $S$-matrix unitarity,
and consider how the BB model can be extended
to significant, real values of the FSA while satisfying the constraints of 
unitarity.

\subsection{S-matrix unitarity in the context of elastic $pp$ scattering}
\label{unitarity}
In this subsection some of the basic equations of quantum scattering theory are recapitulated.
The scattering or $S$ matrix describes how 
a physical system changes in a scattering process. 
The unitarity of the $S$ matrix ensures that the sum of the 
probabilities of all possible outcomes of the scattering process is one.

	The unitarity of the scattering matrix~$S$ is expressed by the equation
	\begin{equation}
	 	SS^{\dagger}=I\,,
		\label{S_matrix_unitarity}
	\end{equation}
	where $I$ is the identity matrix. The decomposition $S=I + iT$, where $T$ is the transition matrix, leads the unitarity relation Eq.~(\ref{S_matrix_unitarity}) to
	\begin{equation}
	 	T - T^{\dagger}=iTT^{\dagger}\,,
		\label{T_matrix_unitarity}
	\end{equation}
	which can be rewritten in the impact parameter $b$ representation as
	\begin{equation}
	 	2\,\text{Im}\,t_{el}(s,b)=|t_{el}(s,b)|^{2} + \tilde\sigma_{inel}(s,b)\,,
		\label{master_equation}
	\end{equation}
	where $s$ is the squared total center-of-mass energy.

	The functions $\tilde\sigma_{inel}(s,b)=d^{2}\sigma_{inel}/d^{2}b$ and $|t_{el}(s,b)|^{2}=d^{2}\sigma_{el}/d^{2}b$ are the inelastic and elastic scattering probabilities
	per unit area, respectively. The elastic amplitude $t_{el}(s,b)$ is defined in the impact parameter space and corresponds to the $\ell$th partial wave amplitude $T_{\ell}(s)$ through the relation $\ell+1/2\leftrightarrow b\sqrt{s}/2$, which
	is valid in the high energy limit, $\sqrt{s}\rightarrow\infty$.

	The unitarity relation~(\ref{master_equation}) is a second order polynomial equation in terms of the (complex) elastic amplitude $t_{el}(s,b)$. If
	one introduces the opacity or eikonal function~\cite{Glauber_lectures,Levin:1998pk,Khoze:2014aca,Ryskin:2012az,Ryskin:2009qf,Martin:2012nm}
	\begin{align}
		t_{el}(s,b)&=i\left[1-e^{-\Omega(s,b)}\right]\,,
 		\label{uBB_ansatz}
	\end{align}
	$\tilde\sigma_{inel}$ can be expressed as
	\begin{align}
	 	\tilde\sigma_{inel}(s,b)&=1-e^{-2\,\text{Re}\,\Omega(s,b)}\, .
 		\label{uBB_ansatz_b}
	\end{align}
	The formula for $t_{el}$ is the so called eikonal form. From Eqs.~(\ref{uBB_ansatz_b}) the real part of the opacity function $\Omega(s,b)$ can be expressed as
	\begin{equation}
	 	\text{Re}\,\Omega(s,b)=-\frac{1}{2}\ln\left[1-\tilde\sigma_{inel}(s,b)\right]\,.
		\label{connection_omega_sigma}
	\end{equation}
	In the original BB model it is assumed that the real part of $t_{el}$ vanishes. In this case Eqs.~(\ref{uBB_ansatz}) and~(\ref{connection_omega_sigma}) imply that
	\begin{align}
	 	t_{el}(s,b)=i\left[1-\sqrt{1-\tilde\sigma_{inel}(s,b)}\right]\,.
		\label{original_BB_model}
	\end{align}
	If the imaginary part $\text{Im } \Omega$ is taken into account in Eq.~(\ref{uBB_ansatz}) the result is
	\begin{align}
	 	&t_{el}(s,b)=i\left[1-e^{-i\,\text{Im}\,\Omega(s,b)}\sqrt{1- \tilde\sigma_{inel}(s,b)}\right]\,,
		\label{unitarized_BB_model}
	\end{align}
	where the concrete parametrization of $\text{Im}\,\Omega(s,b)$ is discussed later.

	To compare the model with data the amplitude~Eq.~(\ref{unitarized_BB_model}) has to be transformed into momentum space
	\begin{align}
		T(s,\Delta)&= \int\limits^{+\infty}_{-\infty}\int\limits^{+\infty}_{-\infty}{e^{i{\vec \Delta} \cdot {\vec b}}{t_{el}(s,b)\text{\rm d}^2b}} \notag\\
		&= 2\pi i\int\limits_0^{\infty}{J_{0}\left(\Delta\cdot b\right)\left[1-e^{-\Omega\left(s,b\right)}\right]b\, {\rm d}b}\,,
		\label{elastic_amplitude}
	\end{align}
	where $b=|{\vec b}|$, $\Delta=|{\vec\Delta}|$ is the transverse momentum and $J_{0}$ is the zero order Bessel-function of the first kind. In the high energy limit, $\sqrt{s}\rightarrow\infty$, $\Delta(t) \simeq \sqrt{-t}$  where $t$ is the squared four-momentum
	transfer. Consequently the elastic differential cross-section can be evaluated as 
	\begin{equation}
		\frac{{\rm d}\sigma}{{\rm d} t}=\frac{1}{4\pi}\left|T\left(s,\Delta\right)\right|^2\,.
		\label{differential_cross_section}
	\end{equation}

	According to the optical theorem the total cross-section is
	\begin{align}
		\sigma_{tot}=2 \left.{\rm Im}\,T(s,\Delta)\right|_{t=0}\,,
		\label{total_cross_section}
	\end{align}
	while the ratio of the real to the imaginary FSA is
	\begin{align}
		\rho=\frac{\text{Re}\,T(s,0)}{\text{Im}\,T(s,0)}\,.
		\label{rho_parameter}
	\end{align}

\subsection{The Bialas-Bzdak model with a unitarily extended amplitude}
\label{s:uBB}
	The original BB model~\cite{Bialas:2006qf} describes the proton as a bound state of a quark and a diquark, where both constituents have to be understood as ``dressed'' objects that effectively include all
	possible virtual gluons and $q\bar{q}$ pairs. The quark and the diquark are characterized with their positions with respect to the proton's center-of-mass using their transverse position vectors $\vec{s}_{q}$ and $\vec{s}_{d}$ in the plane 
	perpendicular to the proton's incident momentum. Hence, the coordinate space~$H$ of the colliding protons is spanned by the vector $h=(\vec{s}_{q},\vec{s}_{d},\vec{s}^{\,\prime}_{q},\vec{s}^{\,\prime}_{d})$ where the
	primed coordinates indicate the coordinates of the second proton. 
	
	The inelastic $pp$ scattering probability $\tilde\sigma_{inel}(s,b)$ in Eq.~(\ref{original_BB_model}) is calculated as an average of ``elementary'' inelastic scattering probabilities $\sigma(h;{\vec b})$ over the 
	coordinate space $H$~\cite{Lipari:2013kta}
	\begin{equation}
	\tilde\sigma_{inel}(b)=\left<\sigma(h;{\vec b})\right>_{H}=\int\limits^{+\infty}_{-\infty}...\int\limits^{+\infty}_{-\infty}{{\rm d}h\,p(h)\cdot\sigma(h;{\vec b})}\,,
	\label{sigma_b_BB}
	\end{equation}
	where the weight function~$p(h)$ is a product of probability distributions
	\begin{equation}
		p(h)=D({\vec s}_q,{\vec s}_d)\cdot D({\vec s}^{\,\prime}_{q},{\vec s}^{\,\prime}_{d})\,.
		\label{probability_distribution}
	\end{equation}
	The $D({\vec s}_q,{\vec s}_d)$ function is a two-dimensional Gaussian, which describes the center-of-mass distribution of the quark and diquark with respect to the
	center-of-mass of the proton
    \begin{equation}
        D\left({\vec s}_q,{\vec s}_d\right)=\frac{1+\lambda^2}{R_{qd}^2\,\pi}e^{-(s_q^2+s_d^2)/R_{qd}^2}\delta^2({\vec s}_d+\lambda{\vec s}_q),\;\lambda=\frac{m_q}{m_d}\,.
	\label{quark_diquark_distribution}
    \end{equation}
	The parameter $R_{qd}$, the standard deviation of the quark and diquark distance, is fitted to the data. Note that the two-dimensional Dirac $\delta$ function preserves the proton's center-of-mass and reduces the dimension of the integral in Eq.~(\ref{sigma_b_BB}) from eight to four.

	Note that the original BB model is realized in two different ways: in one of the cases, the diquark structure is not
	resolved. This is referred to as the $p = (q,d)$ BB model. A more detailed variant is when the diquark is assumed to be a
	composition of two quarks, referred as the $p=(q,(q,q))$.
	Our earlier studies using the $\alpha$BB model
	indicated~\cite{CsorgO:2013kua}, that the $p = (q, d)$ case gives somewhat improved confidence levels
	as compared to the $p=(q,(q,q))$ case. So for the present manuscript we discuss results using the $p = (q,d)$ scenario
	only, however, it is straightforward to extend the investigations to the $p=(q,(q,q))$ case.
 	We have performed these calculations but we do not detail their results here given that they result in fits which are not acceptable at $\sqrt{s}=7$~TeV. 
	To demonstrate that this $p = (q,(q,q))$ model does not work,
	we report only its rather unsatisfactory fit quality when the data analysis is discussed.

	It is assumed that the ``elementary'' inelastic scattering probability $\sigma(h;{\vec b})$ can be factorized in terms of binary collisions among the
	constituents with a Glauber expansion 
    \begin{align}
         \sigma(h;{\vec b})=1-\prod_{a}\prod_{b}\left[1-\sigma_{ab}({\vec b} + {\vec s}^{\,\prime}_{a} - {\vec s}_b )\right]\,,\quad a,b\in\{q,d\}\,,
    \label{Glauber_expansion}
    \end{align}
	where the indices~$a$ and~$b$ can be either quark~$q$ or diquark~$d$.

	The~$\sigma_{ab}\left({\vec s}\right)$ functions describe the probability of binary inelastic collision between quarks and diquarks and are assumed
	to be Gaussian
    \begin{equation}
        \sigma_{ab}\left({\vec s}\right) = A_{ab}e^{-s^2/S_{ab}^2},\;S_{ab}^2=R_a^2+R_b^2,\quad a,b \in \{q,d\}\,,
        \label{inelastic_cross_sections}
    \end{equation}
	where the $R_{q},R_{d}$ and $A_{ab}$ parameters are fitted to the data.

	The inelastic cross-sections of quark, diquark scatterings can be calculated by integrating the probability distributions~Eq.~(\ref{inelastic_cross_sections}) as
    \begin{equation}
    \label{totalinelastic}
        \sigma_{ab,\text{inel}}=\int\limits^{+\infty}_{-\infty}\int\limits^{+\infty}_{-\infty}{\sigma_{ab}\left({\vec s}\right)}\,\text{\rm d}^2s= \pi A_{ab}S_{ab}^2\,.
    \end{equation}
	In order to reduce the number of free parameters,
	% it is assumed 
	we assume that the diquarks are bounded very weakly, hence the ratios of the inelastic cross-sections $\sigma_{ab,\text{inel}}$ satisfy
    \begin{equation}
        \sigma_{qq,\text{inel}}:\sigma_{qd,\text{inel}}:\sigma_{dd,\text{inel}}=1:2:4\,, 
        \label{ratiosforsigma}
    \end{equation}
which means that in the BB model the diquark contains twice as many partons than the quark and also that these quarks and diquarks do not ``shadow'' each other during the scattering process. This assumption is not trivial. 
The $p=(q,(q,q))$ version of the BB model allows for different $\sigma_{qq,\text{inel}}:\sigma_{qd,\text{inel}}:\sigma_{dd,\text{inel}}$ ratios. However,
as it was mentioned before, the $p=(q,(q,q))$ is less favored by the data as compared to the $p = (q,d)$ case presented below.

	Using the inelastic cross-sections~Eq.~(\ref{totalinelastic}) together with the assumption~Eq.~(\ref{ratiosforsigma}) the $A_{qd}$ and $A_{dd}$ parameters
	can be expressed with $A_{qq}$
        \begin{equation}
            A_{qd}=A_{qq}\frac{4R_q^2}{R_q^2+R_d^2}\,,\;A_{dd}=A_{qq}\frac{4R_q^2}{R_d^2}\,.
        \end{equation}
	In this way only five parameters have to be fitted to the data $R_{qd}$, $R_{q}$, $R_{d}$, $\lambda$, and $A_{qq}$. 
%
%Move this part to later text, to the end of the section.	
%
%	In practice we fix $A_{qq}=1$ assuming that head on quark-quark ($qq$)
%	collisions are completely inelastic according to 
%  Eq.~(\ref{inelastic_cross_sections}).
 	
	The last step in the calculation is to perform the Gaussian integrals in the average~Eq.~(\ref{sigma_b_BB}) to obtain a formula for~$\tilde\sigma_{inel}(b)$. The Dirac $\delta$ function
	in~Eq.~(\ref{quark_diquark_distribution}) expresses the protons' diquark position vectors as a function of the quarks position
    \begin{equation}
        {\vec s}_d=-\lambda\,{\vec s}_q,\,\; {\vec s}^{\,\prime}_{d}=-\lambda\,{\vec s}^{\,\prime}_{q}\,.
	\label{Dirac_deltas}
    \end{equation}

	After expanding the products in the Glauber expansion~Eq.~(\ref{Glauber_expansion}) the following sum of contributions is obtained
 	\begin{align}
         \sigma(h;{\vec b})=&\sigma_{qq}+2\cdot\sigma_{qd}+\sigma_{dd}-(2\sigma_{qq} \sigma_{qd}+\sigma_{qd}^{2}+\sigma_{qq}\sigma_{dd}+2\sigma_{qd} \sigma_{dd})\notag\\
		&+(\sigma_{qq}\sigma_{qd}^{2}+2\sigma_{qq}\sigma_{qd}\sigma_{dd}+\sigma_{dd}\sigma_{qd}^{2})-\sigma_{qq}\sigma_{qd}^{2}\sigma_{dd}\,,
	\label{Glauber_expansion_2}
	\end{align}
	where the arguments of the $\sigma_{ab}(\vec{s})$ functions are suppressed to abbreviate the notation.

	The average over H in Eq.~(\ref{sigma_b_BB}) has to be calculated for each term in the above expansion Eq.~(\ref{Glauber_expansion_2}). Take the last, most general, term
	and calculate the average; the remaining terms are simple consequences of it. The result is
	\begin{equation}
		I=\left< -\sigma_{qq}\sigma_{qd}^{2}\sigma_{dd} \right>_{H}=\int\limits^{+\infty}_{-\infty}...\int\limits^{+\infty}_{-\infty}{{\rm d}h\,p(h)\cdot(-\sigma_{qq}\sigma_{qd}^{2}\sigma_{dd}\,)}\,,
		\label{sigma_b_BB_2}
	\end{equation}
	where the $p(h)$ weight function Eq.~(\ref{probability_distribution}) is a product of the quark-diquark distributions, given by Eq.~(\ref{quark_diquark_distribution}). Substitute into this result Eq.~(\ref{sigma_b_BB_2}) the definitions of the quark-diquark distributions Eq.~(\ref{quark_diquark_distribution})
    \begin{align}
		I=-\frac{4v^{2}}{\pi^2}\int\limits^{+\infty}_{-\infty}\int\limits^{+\infty}_{-\infty}{\rm d}^2s_q {\rm d}^2s_q'\,e^{-2v\left(s_q^2+s_q'^2\right)}\prod_{k}\prod_{l}\sigma_{kl}(\vec{b}-\vec{s}_{k}+\vec{s}_{l}^{\,\prime}),\quad k,l \in \{q,d\}\,,
            \label{master_formula_1}
    \end{align}
	where $v=(1+\lambda^{2})/(2\cdot R_{qd}^{2})$ and the integral over the coordinate space $H$ is explicitly written out; it is only four dimensional due to the two Dirac $\delta$ functions in $p(h)$. Using the definitions of the $\sigma_{ab}\left({\vec s}\right)$ functions Eq.~(\ref{inelastic_cross_sections}) and the 
	expression $A=A_{qq}A_{qd}A_{dq}A_{dd}$ the integral Eq.~(\ref{master_formula_1}) can be rewritten, to make all the Gaussian integrals explicit 
    \begin{align}
	I=-\frac{4v^{2}A}{\pi^2}\int\limits^{+\infty}_{-\infty}\int\limits^{+\infty}_{-\infty}{{\rm d}^2s_q {\rm d}^2s_q'\,e^{-2v\left(s_q^2+s_q'^2\right)}
            \prod_{k}\prod_{l}e^{-c_{kl}\left(\vec{b}-\vec{s}_{k}+\vec{s}_{l}^{\,\prime}\right)^2}}\,,
            \label{master_formula_2}
    \end{align}
	where the abbreviations $c_{kl}=S_{kl}^{-2}$ refer to the coefficients in Eq.~(\ref{inelastic_cross_sections}). Finally, the four Gaussian integrals have to be evaluated in our last expression Eq.~(\ref{master_formula_2}),
	which leads to
    \begin{align}
		I=-\frac{4v^{2}A}{B}e^{-b^2\frac{\Gamma}{B}}\,,
            \label{master_formula}
    \end{align}
	where
    \begin{align}
        B&=C_{qd,dq}\left(v+c_{qq} + \lambda^2 c_{dd}\right)+\left(1-\lambda\right)^2 D_{qd,dq}\,,\notag\\
        \Gamma&=C_{qd,dq}D_{qq,dd} + C_{qq,dd}D_{qd,dq}\,,
            \label{master_formula1}
    \end{align}
	and
    \begin{align}
	C_{kl,mn}&=4v + \left(1+\lambda\right)^2\left(c_{kl}+c_{mn}\right)\,,\notag\\
	D_{kl,mn}&=v \left(c_{kl}+c_{mn}\right)+\left(1+\lambda\right)^2c_{kl}c_{mn}\,.
	            \label{master_formula2}
    \end{align} 
	
	Each term in~Eq.~(\ref{sigma_b_BB}) can be obtained from the master formula Eq.~(\ref{master_formula}), by setting one or more coefficients to
	zero,~$c_{kl}=0$ and the corresponding amplitude to one,~$A_{kl}=1$.

Up to now, we have evaluated the real part of the opacity or eikonal function $\Omega(s,b)$, that is determined by the inelastic scattering probability per unit area according to Eq.~(\ref{connection_omega_sigma}).
Now we also have to specify the imaginary part of the complex opacity function, that determines the real part of the FSA. Here several model assumptions are possible, but from the analysis of the ISR data and the first studies of the 7 TeV TOTEM data at LHC we learned, that the real part of the FSA is perturbatively small at ISR energies, it becomes non-perturbative at LHC but the scattering is still dominated by the imaginary part of the scattering amplitude.

We have studied several possible choices. One possibility is to introduce the imaginary part of the opacity function so that it is proportional to the probability of inelastic scatterings, which is known to be a decreasing function of the impact parameter $b$. A possible interpretation of this assumption may be that the inelastic collisions arising from non-collinear elastic collisions of quarks and diquarks follow the same spatial distributions as the inelastic collisions of the same constituents
	\begin{equation}
		\text{Im}\,\Omega(s,b)=-\alpha\cdot\tilde\sigma_{inel}(s,b)\,,
		\label{omega_b_with_sigma_b}
	\end{equation}
where $\alpha$ is a real number, corresponding to the shape parameter
of the differential cross-section of elastic $pp$ scattering.
\footnote{
It is interesting to note a similarity of this concept with the so called ``Pomeron elasticity'', which was introduced independently for quark-diquark models  in Refs.~\cite{Grichine:2012ry} and \cite{Grichine:2014wea}. Pomeron elasticity 
has a similar shape modifying role in the dip region: it was introduced  to increase the real part of FSA by modifying the standard Pomeron trajectory in elastic quark-quark, quark-diquark and diquark-diquark scattering, that contributes significantly in particular close to
the dip region of $pp$ elastic scattering.}

The above proportionality Eq.~(\ref{omega_b_with_sigma_b}) between~$\text{Im}\,\Omega(s,b)$ and $\tilde\sigma_{inel}(s,b)$ provided the best fits from among the relations that we have tried
but it is far from being a unique possibility for an ansatz. Note that the $\alpha = 0$ case corresponds to the $p=(q,d)$ version of the original BB model of Ref.~\cite{Bialas:2006qf}, where the FSA has
a vanishing real part. 
	
In the $\alpha < 0.1$ perturbative limit
the $\alpha$BB model of Ref.~\cite{CsorgO:2013kua} can also be obtained 
as follows.
	The $\alpha$BB model can be defined by the relation
	\begin{align}
		\text{Im}\,\Omega(s,b) &= \frac{\alpha\cdot\tilde\sigma_{inel}(s,b)}{\tilde\sigma_{inel}(s,b)-2}\,.
	\label{omega_aBB}
	\end{align}
	This definition is equivalent to the original form of the definition of the $\alpha$BB model in Ref.~\cite{CsorgO:2013kua} where $\sigma_{inel}(s,b)$ in Eq.
	(\ref{original_BB_model}) was allowed to have a small imaginary part, in the form of 
	$\sigma_{inel}(s,b)\, \rightarrow \, (1+ i \alpha) \sigma_{inel}(s,b)$.  This condition was satisfied for fits at ISR energies,
	with $\alpha \approx 0.01$, however, at LHC energies the $\alpha$BB model
	did not describe the TOTEM data in a satisfactory manner~\cite{CsorgO:2013kua}.

We have also investigated the assumption that the real and the imaginary parts of the opacity function are proportional to one another
	\begin{equation}
		\text{Im}\,\Omega(s,b)=-\alpha\cdot\text{Re}\,\Omega(s,b)\,.
		\label{omega_real_b_omega_imag_b}
	\end{equation}

However, as the results using Eq.~(\ref{omega_real_b_omega_imag_b}) were less favorable than the results obtained with Eq.~(\ref{omega_b_with_sigma_b}), we do the data analysis part,  described in the next section, using Eq.~(\ref{omega_b_with_sigma_b}).

We mention this possibility to highlight that here some phenomenological assumptions are necessary as the ReBB model does allow for a broad range of possibilities for the choice of the imaginary part of the opacity function.

In this way, the ReBB model is fully defined, and at a given colliding energy only six parameters determine the differential~(\ref{differential_cross_section}) and total cross-sections~(\ref{total_cross_section}) and	also the $\rho$ parameter, defined with Eq.~(\ref{rho_parameter}). The parameters that have to be fitted to the data include the three scale parameters, $R_{q}$, $R_{d}$, $R_{qd}$,  that fix the geometry of the $pp$ collisions, as well as the three additional parameters $\alpha$, $\lambda$ and $A_{qq}$.  Two of the latter three can be fixed:   $\lambda = 0.5$ if the diquark is very weakly bound, so that its mass is twice as large as that of the valence quark. 
In practice we also fix  $A_{qq} = 1$, assuming that head-on $qq$ collisions are inelastic with a probability of 1, according to Eq.~(\ref{inelastic_cross_sections}).

Thus in the present data analysis two out of six parameters are fixed and only four parameters are fitted to the data at each $\sqrt{s}$: the three scale parameters $R_{q}$, $R_{d}$ and  $R_{qd}$,  as well as the shape parameter $\alpha$. As we shall see, the shape parameter $\alpha$ will play a key role  when describing the shape of the dip of the differential cross-sections of elastic $pp$ scatterings at LHC energies.

\section{Fit method and results}
\label{sec:fitdescription}

	The $pp$ elastic differential cross-section data measured by the TOTEM experiment at $7$~TeV is a compilation of two subsequent measurements~\cite{Antchev:2011vs,Antchev:2011zz,Antchev:2013gaa}.
	The squared four-momentum transfer value $t_{sep}=-0.375$~GeV$^{2}$ separates the two data sets.\footnote{The squared four-momentum transfer value $t_{sep}$ separates the bin centers at the common boundary,
	the two bins actually overlap~\cite{Antchev:2011vs,Antchev:2011zz,Antchev:2013gaa}.} Note, that the two datasets were taken with two very different settings of the machine optics of the LHC accelerator.

	The ReBB model, defined with Eqs.~(\ref{unitarized_BB_model}) and~(\ref{omega_b_with_sigma_b}), was fitted to the data at ISR energies and at LHC energy of $\sqrt{s}=7$~TeV. The relation between the imaginary part of $\Omega(s,b)$ and $\alpha$ is defined with Eq.~(\ref{omega_b_with_sigma_b}).
	In agreement with our previous investigations the $A_{qq}=1$ and $\lambda=\frac{1}{2}$ parameters can be kept constant, which reduces the number of free parameters to four $R_{qd},\,R_{q},\,R_{d}$ and $\alpha$.
	The case when the parameters $A_{qq}$ and $\lambda$ are also free is summarized at the end of this section.

	In the course of the minimization of the ReBB model we take into account the uncertainty of the overall scale factor of the measured data. The fitted $\chi^{2}$ function to the
	data points is
	\begin{equation}
        	\chi^{2}=\sum_{i=1}^{N}\frac{\left({\rm d}\sigma_{i}/{{\rm d}t}-\gamma\cdot{{\rm d}\sigma_{th,i}}/{{\rm d}t}\right)^{2}}{\sigma_{i}^{2}}+\frac{\left(\gamma_{0} - \gamma\right)^{2}}{\sigma_{lumi}^{2}}\, ,
        	\label{differential_cross_section_gamma}
    	\end{equation}
    	where $N$ is the number of fitted data points, ${\rm d}\sigma_{i}/{{\rm d}t}$ is the $i$th measured differential cross-section data point and ${{\rm d}\sigma_{th,i}}/{{\rm d}t}$ is the corresponding theoretical value at the
	$i$th data point calculated from the ReBB model. The value $\sigma_{i}$ is the mere statistical uncertainty of the $i$th data point.
	We set $\sigma_{lumi}$ to a conservative value of $5~\%$ at each ISR energy\cite{Amaldi:1979kd}. In case of the TOTEM data sets it is set to $4~\%$, which is determined by the uncertainty of the CMS luminosity\cite{Antchev:2011zz,Antchev:2013gaa}.
	Parameter $\gamma$ is
	the additional normalization parameter to be minimized and 
	$\gamma_{0}\equiv 1$ stands for the value of luminosity scaled to unity, while $\sigma_{lumi}$ is the relative luminosity uncertainty reported by the measurement.
	
	The sum in Eq.~(\ref{differential_cross_section_gamma}) runs over the fitted data points, and the additional term takes into account the contribution of the luminosity uncertainty\cite{Demortier}.

\begin{figure}[H]
\centering
\includegraphics[width=0.8\linewidth]{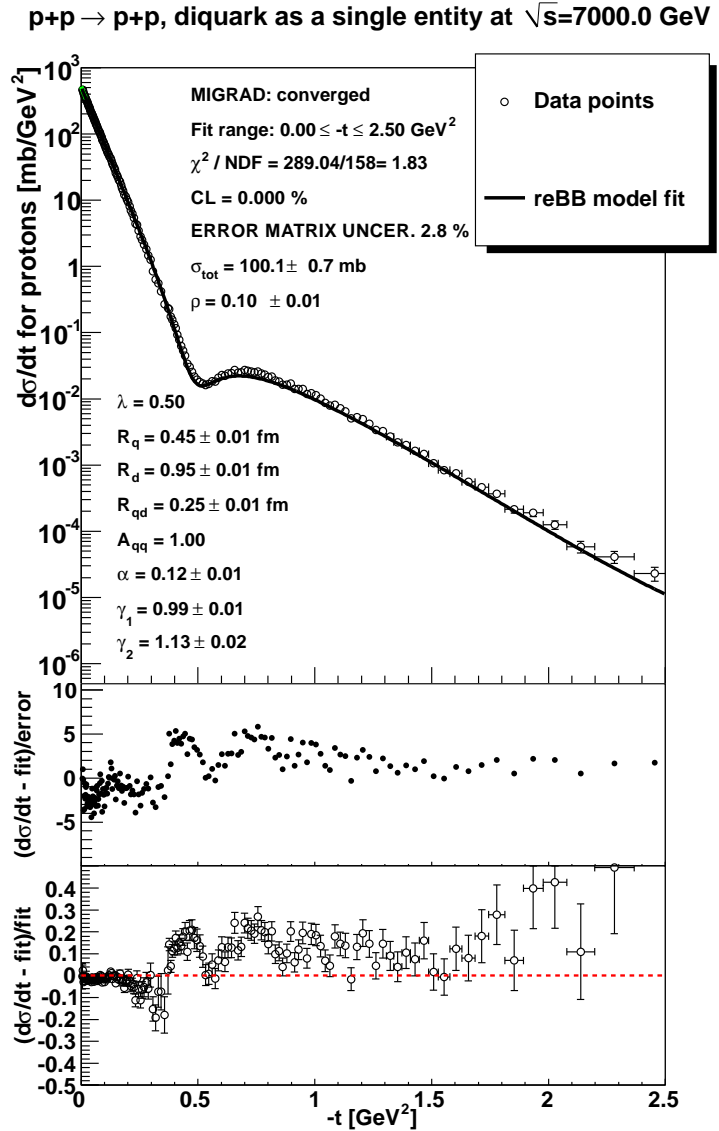}
\caption{
The fit of the ReBB model in the whole $0<|t|< 2.5 $ GeV$^2$ range at $\sqrt{s}=7$~TeV. The fit uses the statistical errors of the data points and the luminosity error of the systematic uncertainty according to Eq.~(\ref{differential_cross_section_gamma}).
Although the fit quality is not satisfactory, CL $\ll$ 0.1\%, the fit looks good by eye.
The fitted parameters are shown in the left bottom corner, parameters without errors were fixed during the MINUIT optimization. The total cross-section~$\sigma_{tot}$
and the parameter $\rho$ are derived quantities according to Eqs.~(\ref{total_cross_section}) and~(\ref{rho_parameter}), respectively. Parameter values are rounded up to two decimal digits.
The uncertainty of the MINUIT error matrix after the fit is 2.8~\%.}

\label{BBm_model_fit_results_full_1}
\end{figure}

	First we attempted to fit the ReBB model in the $0 < |t| < 2.5$
GeV$^{2}$ range, fitting simultaneously both the low-$|t|$ TOTEM measurement of Ref.~\cite{Antchev:2013gaa} and the one containing the dip
region\cite{Antchev:2011zz}, see~Fig.~\ref{BBm_model_fit_results_full_1}. Note, that in this particular fit two normalization parameter were used: $\gamma_{1}$ below and~$\gamma_{2}$ above $|t_{sep}|$, since
the mentioned two TOTEM measurements are independent. This fit provides 
$\chi^{2}/NDF=289.04/158=1.83$
and $CL= 9\times 10^{-9} \ll 0.1\%\,$, which is statistically not an acceptably
good fit quality, although, as indicated in Fig.~\ref{BBm_model_fit_results_full_1}, the fit
looks reasonably good by eye. So, unfortunately, we could not get a
unified and statistically acceptable description of the differential
cross-section of elastic $pp$ scattering in the whole measured $t$
interval in the framework of the unitarily extended Bialas-Bzdak model.

It is important to note, that we determined the fit quality using the 
statistical and the luminosity uncertainty only, where the latter is a $t$-independent systematic
uncertainty. According to the original TOTEM
publications~\cite{Antchev:2011zz,Antchev:2013gaa}, the systematic
uncertainties of the two TOTEM data sets are very different due to different
data taking conditions, especially due to the different LHC machine optics. The
$t$-dependent part of the systematic errors  allow the data points, as a
function of $|t|$, to be slightly moved in a correlated and $t$
dependent way, which could, in principle, improve our fit quality.
	
\begin{figure}[H]
\centering
\includegraphics[width=0.8\linewidth]{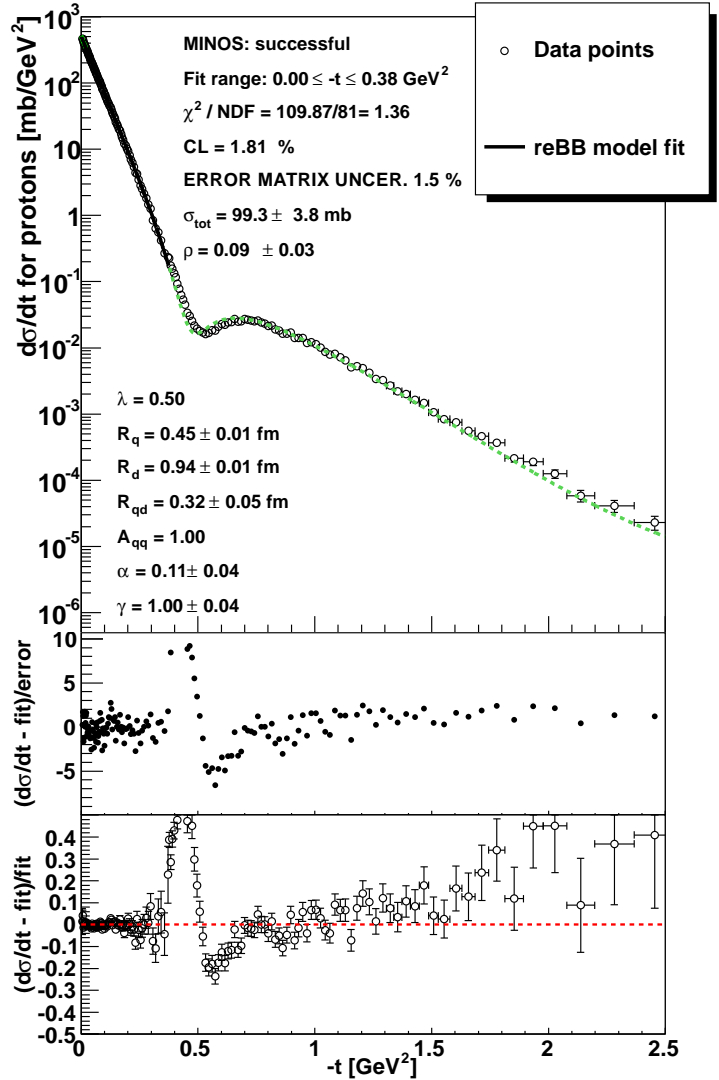}
\caption{The fit of the ReBB model in the $0<|t|<|t_{sep}|$ range at $\sqrt{s}=7$~TeV.
The fit quality is satisfactory, CL $ > 0.1 \%.$
 The fit uses the statistical errors of the data points and the luminosity error of the systematic uncertainty according to Eq.~(\ref{differential_cross_section_gamma}). The
fitted curve is shown with solid line, its extrapolation above $|t_{sep}|$ is indicated with a dashed line. The extrapolated curve remains close to the data points, following the measured differential cross-sections well even far away from the fitted region. 
The fitted parameters are shown in the left bottom corner, rounded to two valuable decimal digits. The parameters without errors were fixed in the minimization. The total cross-section~$\sigma_{tot}$ and the parameter $\rho$ are derived quantities according to Eqs.~(\ref{total_cross_section}) and~(\ref{rho_parameter}), respectively.
The uncertainty of the MINUIT error matrix after the fit is 1.5~\%.}

\label{BBm_model_fit_results_full}
\end{figure}

\begin{figure}[H]
\centering
\includegraphics[width=0.8\linewidth]{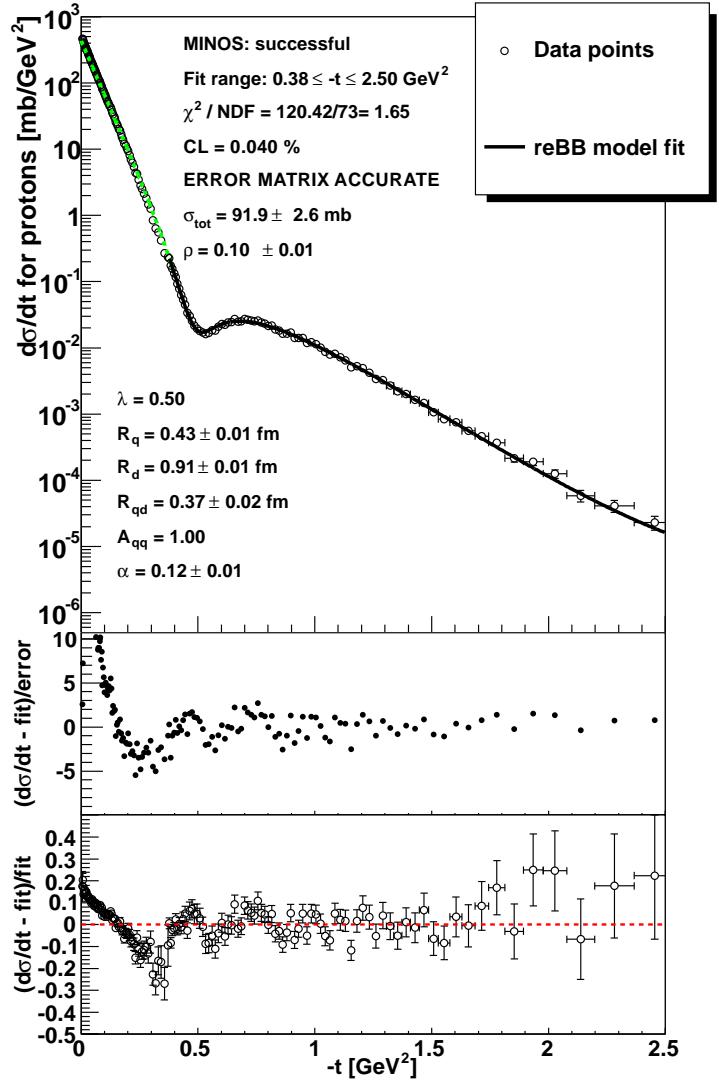}
\caption{The same as Fig.~\ref{BBm_model_fit_results_full}, but the fit is performed in the $|t_{sep}|<|t|<2.5$~GeV$^{2}$ range. The fit uses the statistical errors of the data points and we present the results for the $\gamma = 1 $ fixed case. The fitted curve is shown with solid line,
its extrapolation is indicated with a dashed line. Parameter values are rounded up to two valuable decimal digits.
Note that when the curve is extrapolated to the low-$|t|$ region,  the extrapolated curve again follows the measured differential cross-section remarkably well even far away the fit region: the ReBB model fit is remarkably stable over the whole $|t|$-range.}
\label{BBm_model_fit_results_fulla_cst}
\end{figure}

	However, this part of the systematics is rather difficult to handle correctly in the present analysis. So we decided to analyze the two TOTEM data sets separately. 
If the {\it separated} fits to $\sqrt{s}=7$~TeV elastic differential cross-section $d\sigma/dt$ data are evaluated, below and above the separation $|t_{sep}|$, quality results can be obtained, which are shown in Figs.~\ref{BBm_model_fit_results_full}
and~\ref{BBm_model_fit_results_fulla_cst}, and are summarized in Table~\ref{table:fit_parameters}.
As detailed below, this strategy leads to reasonable fit qualities (CL = 1.8~\%, statistically acceptable
	fit in the cone region and CL = 0.04~\%, statistically marginal fit in the dip region), with
	a remarkable stability of fit parameters.\footnote{The  $\alpha$BB version of the ReBB model~\cite{CsorgO:2013kua} has already been fitted to ISR data but only in the restricted $0.36<|t|<2.5$~GeV$^{2}$ range in order to be consistent with the $t$-range of the available TOTEM data set of that time~\cite{Antchev:2011zz}, while in this work, the low $|t|$ data are also included at each energies. The more limited fit
range explains the seemingly better $\chi^{2}/NDF$ and CL values reported in our earlier publication Ref.~\cite{CsorgO:2013kua} using the $\alpha$BB model.}

\begin{table}[h]\small
\centering
\begin{tabular}{|c|c|c|c|c|c|c|c|} \hline
$\sqrt{s}$ [GeV] & 23.5                 & 30.7			& 52.8			& 62.5			& 	\multicolumn{2}{|c|}{7000}			\\ \hline\hline
$|t|$ [GeV$^{2}$] & \multicolumn{4}{|c|}{$(0,2.5)$}							 	& $(0,|t_{sep}|$) & $(|t_{sep}|,2.5)$ \\ \hline
$\chi^{2}/NDF$	&  124.7/101            &   95.6/46          	&     96.1/47      	&    76.2/46		& 109.9/81  		 & 	120.4/73    \\ \hline
CL [\%] 	& 5.5                   & $2\times10^{-3}$	& $3\times10^{-3}$      & 0.3    		& 1.8		&  $4\times10^{-2}$	 \\ \hline
$R_{q}$ [$fm$] 	& 0.27$\pm$0.01   	& 0.28$\pm$0.01     	& 0.28$\pm$0.01     	& 0.28$\pm$0.01     	& 0.45$\pm$0.01 & 	0.43$\pm$0.01	\\ \hline
$R_{d}$ [$fm$] 	& 0.72$\pm$0.01   	& 0.74$\pm$0.01     	& 0.74$\pm$0.01     	& 0.75$\pm$0.01     	& 0.94$\pm$0.01 & 	0.91$\pm$0.01	\\ \hline
$R_{qd}$ [$fm$] & 0.30$\pm$0.01   	& 0.29$\pm$0.01     	& 0.31$\pm$0.01     	& 0.32$\pm$0.01     	& 0.32$\pm$0.05 & 	0.37$\pm$0.02	\\ \hline
$\alpha$ 	& 0.03$\pm$0.01   	& 0.02$\pm$0.01     	& 0.04$\pm$0.01     	& 0.04$\pm$0.01     	& 0.11$\pm$0.04	&  0.12$\pm$0.01			\\ \hline
$\gamma$ 	& 1.01$\pm$0.05     	& 0.98$\pm$0.05       	& 0.90$\pm$0.06       	& 0.97$\pm$0.05     	& 1.00$\pm$0.04		& 1.00 (fixed)	\\ \hline
\end{tabular}
\caption{The values of the fitted ReBB model parameters from ISR to LHC energies. At the 7 TeV LHC energy, the $pp$ elastic $d\sigma/dt$ data measured by the TOTEM experiment is a composition
of two subsequent measurements, which are separated at $t_{sep}=-0.375$ GeV$^2$. The errors and the values are rounded up to two valuable decimal digits.
}
		\label{table:fit_parameters}
\end{table}

It is quite remarkable, that although the low-$|t|$ fit ends at $t_{sep}=-0.375$ and the dip position is at significantly larger values of $|t|$, still the fit when extrapolated to the dip and the Orear region after the dip reproduces the data very well~\cite{Orear:1964zz}.

	The calculated total cross-section of the low-$|t|$ fit is $\sigma_{tot}=99.3\pm 3.8$~mb, where the uncertainty is the propagated uncertainty of the fit parameters, including that of the shape parameter that can be relatively badly determined in the cone region. This result nevertheless agrees very well with the value
	$\sigma_{tot}=98.0\pm2.5$~mb measured by the TOTEM experiment at $\sqrt{s}=7$~TeV~\cite{Antchev:2013iaa} in a luminosity independent way,
	and the calculated value for the parameter $\rho = 0.09 \pm 0.03$ is also reasonable,
	as within its errors it is consistent with the measured value of $\rho= 0.145 \pm0.091$ as reported by the TOTEM experiment~\cite{Antchev:2013iaa}.
 Note, that the uncertainty of the value of our $\rho$ parameter is the propagated uncertainty of the ReBB model fit parameters as it is given in Fig.~\ref{BBm_model_fit_results_full}, consequently, it contains the effects of propagated statistical and luminosity uncertainties only.

On the other hand, if we look at the fit to the dip region, with $\gamma=1$ fixed, we also see a remarkable stability of the shape of the differential cross-section at low values of $|t|$ that still yield reasonable values for the total cross-section ($\sigma_{tot} = 91.9 \pm 2.6$ mb) and similarly reasonable values for the parameter $\rho = 0.10 \pm 0.01$.
The stability and consistency of the model description is visible also in Figs.~\ref{BBm_model_fit_results_full} and \ref{BBm_model_fit_results_fulla_cst}.

\begin{figure}[H]
	\centering
\includegraphics[width=0.8\linewidth]{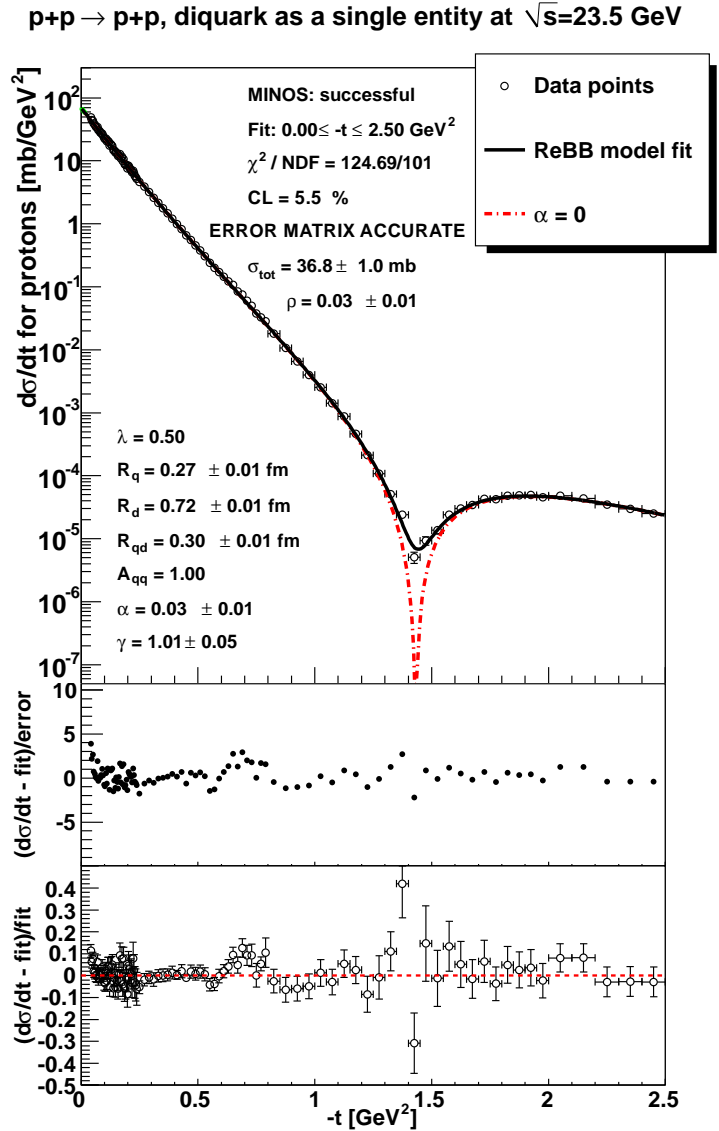}
\caption{The fit of the ReBB model at $\sqrt{s}=23.5$~GeV in the $0<|t|<2.5$~GeV$^{2}$ squared four-momentum transfer $|t|$ range. The fit uses the statistical errors of the data points and the luminosity error
of the systematic uncertainty according to Eq.~(\ref{differential_cross_section_gamma}). Parameter values are rounded up to two valuable decimal digits.
}
\label{BBm_model_fit_results_fullb}
\end{figure}

For the sake of completeness, we present also one of our fits at the ISR energy of  $\sqrt{s}=23.5$~GeV, as indicated in Fig.~\ref{BBm_model_fit_results_fullb}~\cite{Antchev:2011vs,Antchev:2013gaa,Nagy:1978iw,Amaldi:1979kd}. The parameters of
the best fits and the parameters' errors at each analyzed ISR energy are summarized in Table~\ref{table:fit_parameters}.
 
If the two parameters $A_{qq}$ and $\lambda$ are released and
included into the set of fitted parameters of the ReBB model, the fit quality
improves at each analyzed energy from the point of view of
mathematical statistics. In case of
$\sqrt{s}=30.7$~GeV the improvement is quite significant as the confidence 
level of the fit reaches  $CL = 8\,\%$ instead of
$2\times10^{-3}\,\%$. However, the two new fitted parameters introduce more
correlations, which lead to large fit uncertainties, 
and the parameters $\lambda$ and $A_{qq}$ within errors
remain in the range of their values that were fixed for the fits reported
in Table~\ref{table:fit_parameters}, however,
due to the correlations between the fit parameters when $A_{qq}$ and $\lambda$
are released, the fit parameters fluctuate more when evaluated as a function of  ~$\sqrt{s}$, consequently their trend is more difficult to determine. Therefore, 
in order to determine the excitation function of the model parameters,
we utilized the  results of the ReBB fit results as listed 
in Table~\ref{table:fit_parameters},  where these two parameters $\lambda=1/2 $ and 
$A_{qq} = 1$ are fixed.

Also note that if $\text{Im}\,\Omega(s,b)$ is defined to be proportional to
$\text{Re}\,\Omega(s,b)$, according to Eq.~(\ref{omega_real_b_omega_imag_b}),
the MINUIT fit result of $\chi^{2}/NDF=504.9/159=3.2$ is
obtained at $\sqrt{s}=7$~TeV in the $0 < |t| < 2.5$ GeV$^{2}$
range, which is disfavored as compared to fits with
Eq.~(\ref{omega_b_with_sigma_b}), see also Fig.~\ref{BBm_model_fit_results_full_1}.

In our introduction we shortly mentioned the $p=(q,(q,q))$ version of the ReBB
model, when the diquark is assumed to be a composition of two
quarks~\cite{CsorgO:2013kua}.  At $\sqrt{s}=7$~TeV in the $0 <
|t| < 2.5$ GeV$^{2}$ range this scenario provides a fit result with
$\chi^{2}/NDF=15509/159 \approx 97.5$, which means that the $p=(q,(q,q))$ ReBB
version can be rejected. The failure of this version is basically due
the wrong shape of the differential cross-section: the second diffractive
minimum appears too close to the first one.

\section{Discussion}
\label{sec:discussion}
\subsection{Shadow profile functions and saturation}
\label{sec:shadow_profile}

\begin{figure}[h]
\centering
\includegraphics[width=0.48\linewidth]{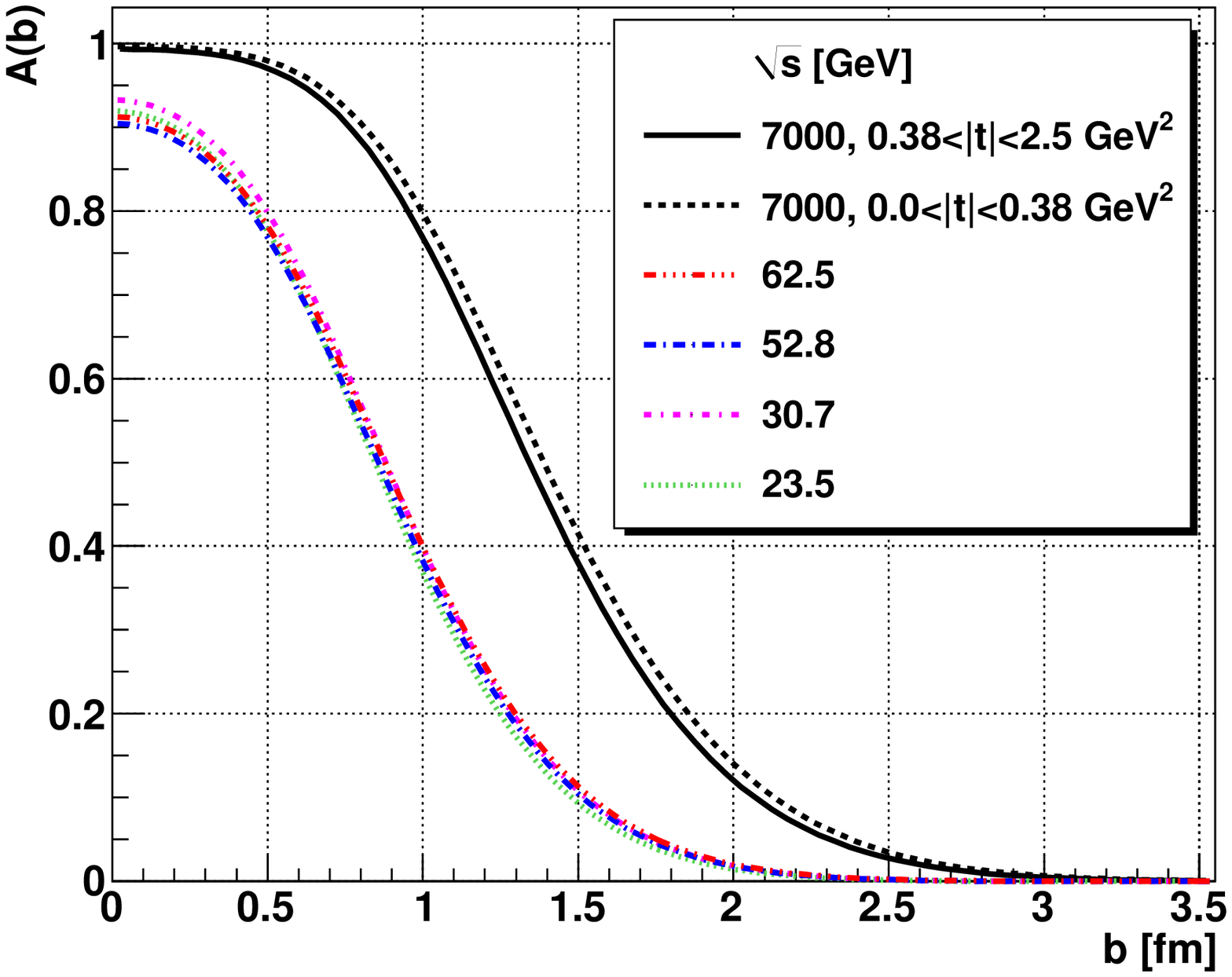}
\includegraphics[width=0.48\linewidth]{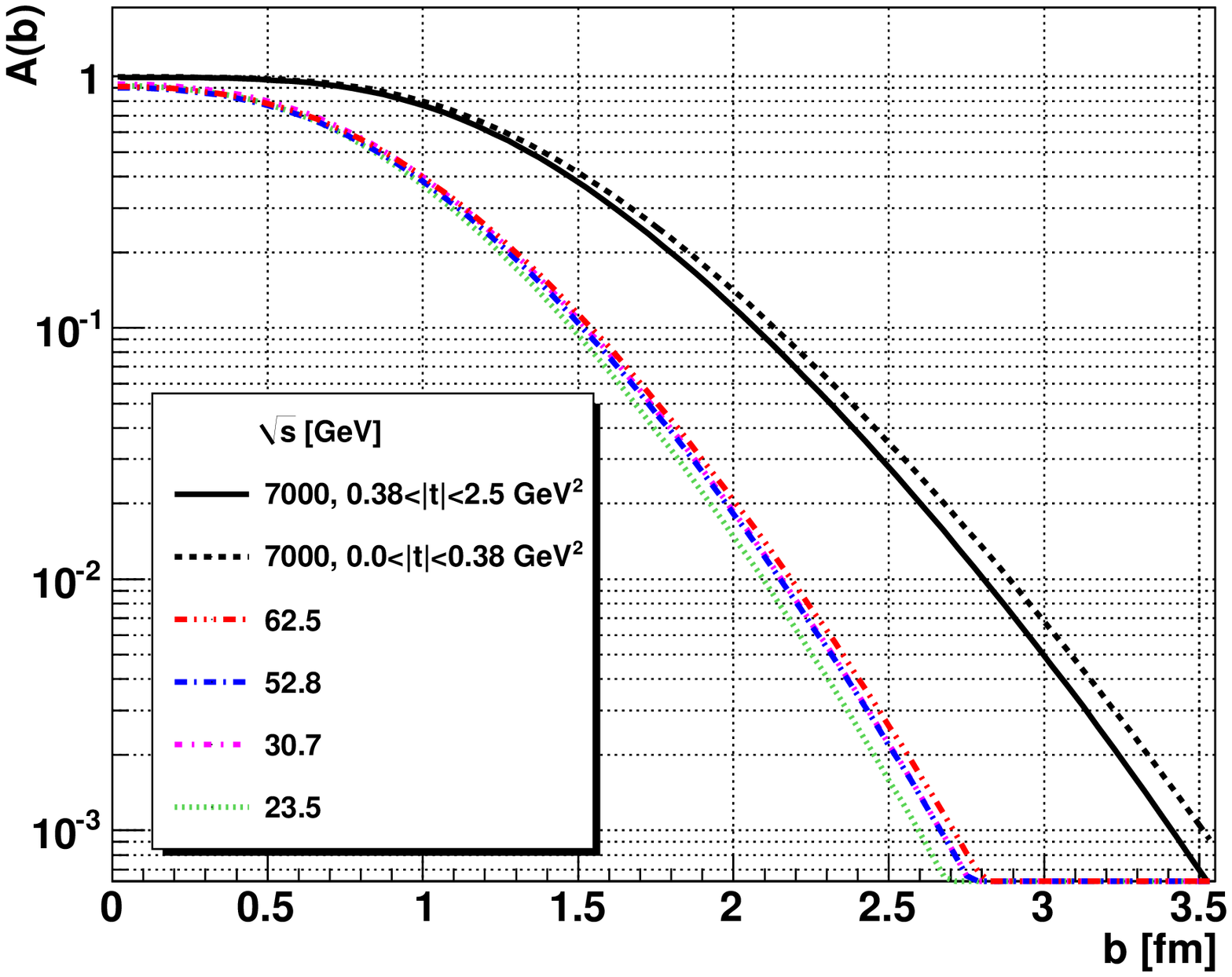}
\caption{The shadow profile functions $A(b)$ indicate a saturation
effect at LHC, while at ISR energies a Gaussian shape can be observed. Note
that the 7 TeV (black dashed) curve is based on the statistically acceptable fit
result in the $0<|t|<0.38$~GeV$^{2}$ range. The distributions' edge shows
approximately the same width at each energy, corresponding to a constant
``skin-width'' of the proton.} 
\label{BBm_model_fit_results_shadow}
\end{figure}

The fits, from which the model parameters were determined, also permit us to
evaluate the shadow profile function \begin{equation}
A(s,b)=1-\left|\exp\left[-\Omega(s,b)\right]\right|^{2}\,.  \end{equation} The
obtained curves to $A(b)$ are shown in Fig.~\ref{BBm_model_fit_results_shadow}.
The shadow profile functions at ISR energies exhibit a Gaussian like shape,
which smoothly change with the center-of-mass energy $\sqrt{s}$.
In this case, the $A(s,b=0)<1$ value indicates that the protons
are not completely ``black" at the ISR energies, even at their centre they do
not scatter with the maximum possible probability per unit area. At the LHC energy
of $\sqrt{s} = 7$~TeV something new appears: the innermost part of the
distribution shows a saturation, which means that around $b=0$
the shadow profile function becomes almost flat and stays close to
$A(b)\approx 1$. Consequently, the shape of the shadow profile function $A(b)$
becomes non-Gaussian and somewhat ``distorted'' with respect to the shapes
found at ISR.  At the same time, the width of the edge of the shadow profile
function $A(b)$, which can be visualized as the proton's ``skin-width'', remains
approximately independent of the center-of-mass energy $\sqrt{s}$.

\subsection{Non-exponential behavior of $d\sigma_{el}/dt$}
\label{sec:non_exponential}

To compare the obtained low-$|t|$ ReBB fit 
of Fig.~\ref{BBm_model_fit_results_full} with a {\it purely} exponential
distribution the following exponential parametrization is used
\begin{equation}
	\frac{\rm{d}\sigma_{el}}{\rm{d}t} = 
	\left.\frac{\rm{d}\sigma_{el}}{\rm{d}t}\right|_{t=0}\cdot e^{-B\cdot|t|}\,,
	\label{exp_distribution_from_Jan}
\end{equation}
where $\left.\rm{d}\sigma_{el}/\rm{d}t\right|_{t=0}=506.4$~mb/GeV$^{2}$ 
and the slope parameter of $B=19.89$~GeV$^{-2}$ is applied,
according to the TOTEM paper of Ref.~\cite{Antchev:2013gaa}.

\begin{figure}[h] \centering
\includegraphics[width=0.8\linewidth]{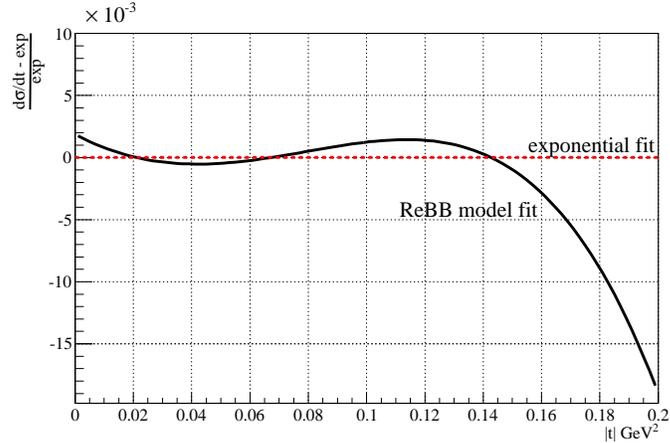}
\caption{ 
The ReBB model fit result, shown in
Fig.~\ref{BBm_model_fit_results_full}, with respect to the exponential fit of
Eq.~(\ref{exp_distribution_from_Jan}).  In the plot only the
$0.0\le|t|\le0.2$~GeV$^{2}$ range is shown, but the ReBB model is fitted to 7
TeV TOTEM data in the low-$t$ interval of $(0,t_{sep})$. The ReBB fit result
indicates a significant deviation from the simple exponential at low-$|t|$
values.} 

\label{BBm_model_fit_results_non_exponential} 
\end{figure}

	The result, shown in~Fig.~\ref{BBm_model_fit_results_non_exponential},
indicates a clear non-exponential behavior of the elastic differential
cross-section in the $0.0\le|t|\le0.2$~GeV$^{2}$ range at $\sqrt{s}=$7 TeV.

A change of the slope parameter $B(t, s) = ({\rm d}/{\rm d}t)\ln ({\rm d}\sigma/{\rm d}t)$ 
around $-t \approx 0.10$ GeV$^2$ was reported already in the year
1972 in elastic $pp$ collisions at the ISR energy 
range of 21.3 $ < \sqrt{s} < $ 54 GeV and the 
0.02~GeV$^2<-t<$~0.40~GeV$^2$ squared four-momentum transfer range. 
A very similar structure, a deviation from an 
exponential behavior was also reported  as early as in 1984 
in the analysis of proton-antiproton elastic scattering at 
546 GeV by Glauber and Velasco~\cite{Glauber:1984su}.
The first preliminary results on such a non-exponential behavior
at the CERN LHC energy of 8 TeV were made public recently in $pp$
elastic scattering by the TOTEM experiment at various conference presentations
during 2014~\cite{Simone:WPCF2014} and were already interpreted
in the theoretical work of Ref.~\cite{Jenkovszky:2014yea},
as the consequence of two-pion exchange and $t$-channel unitarity, while
Ref.~\cite{Khoze:2014nia}, related this $t$ dependence 
of the slope parameter $B$ to pion loop contribution
to the Pomeron trajectory in two-channel eikonal models.

\section{Extrapolation to future LHC energies and beyond}
\label{sec:excitation}

\begin{figure}[h] \centering
\includegraphics[width=0.495\linewidth]{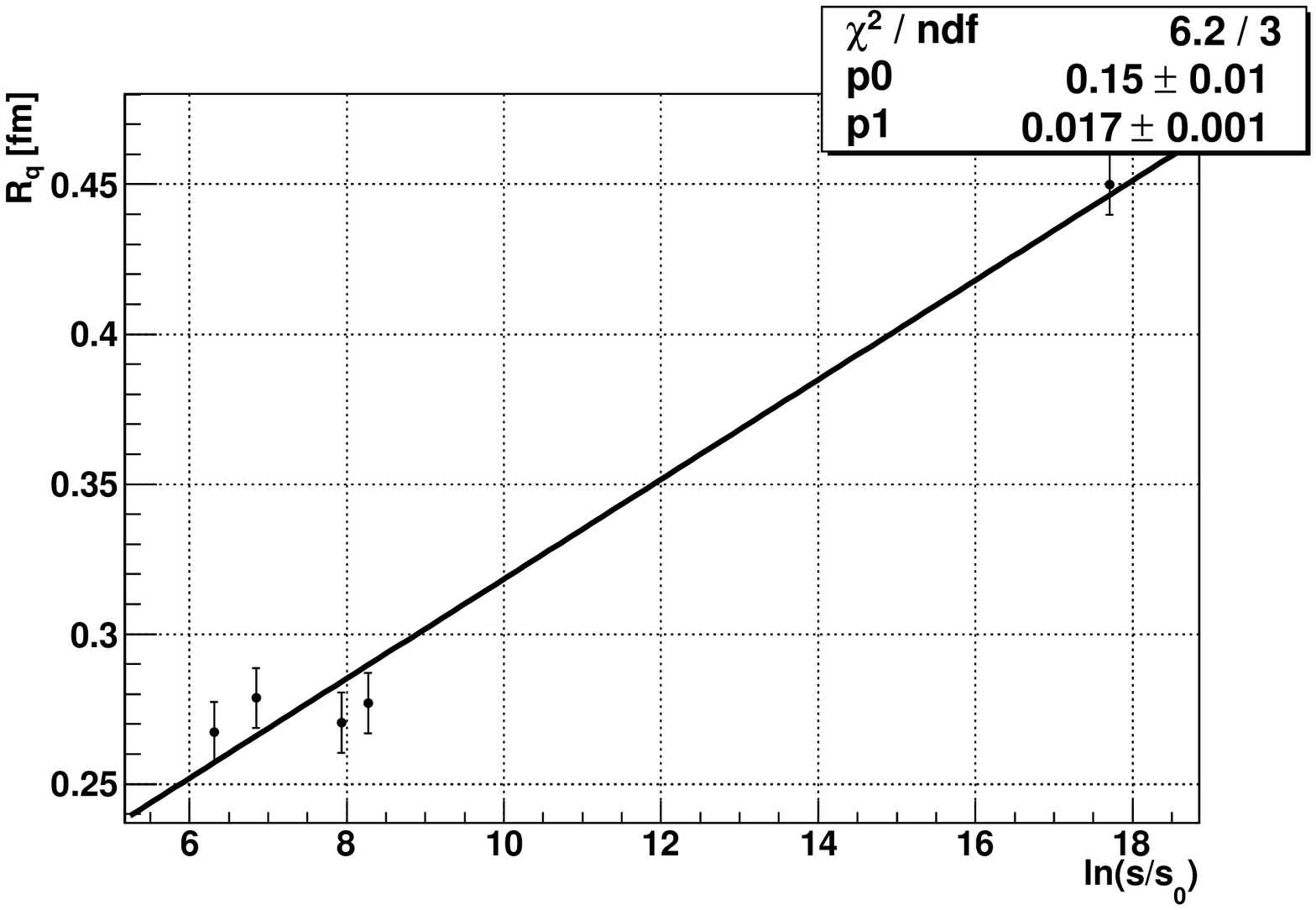}
\includegraphics[width=0.495\linewidth]{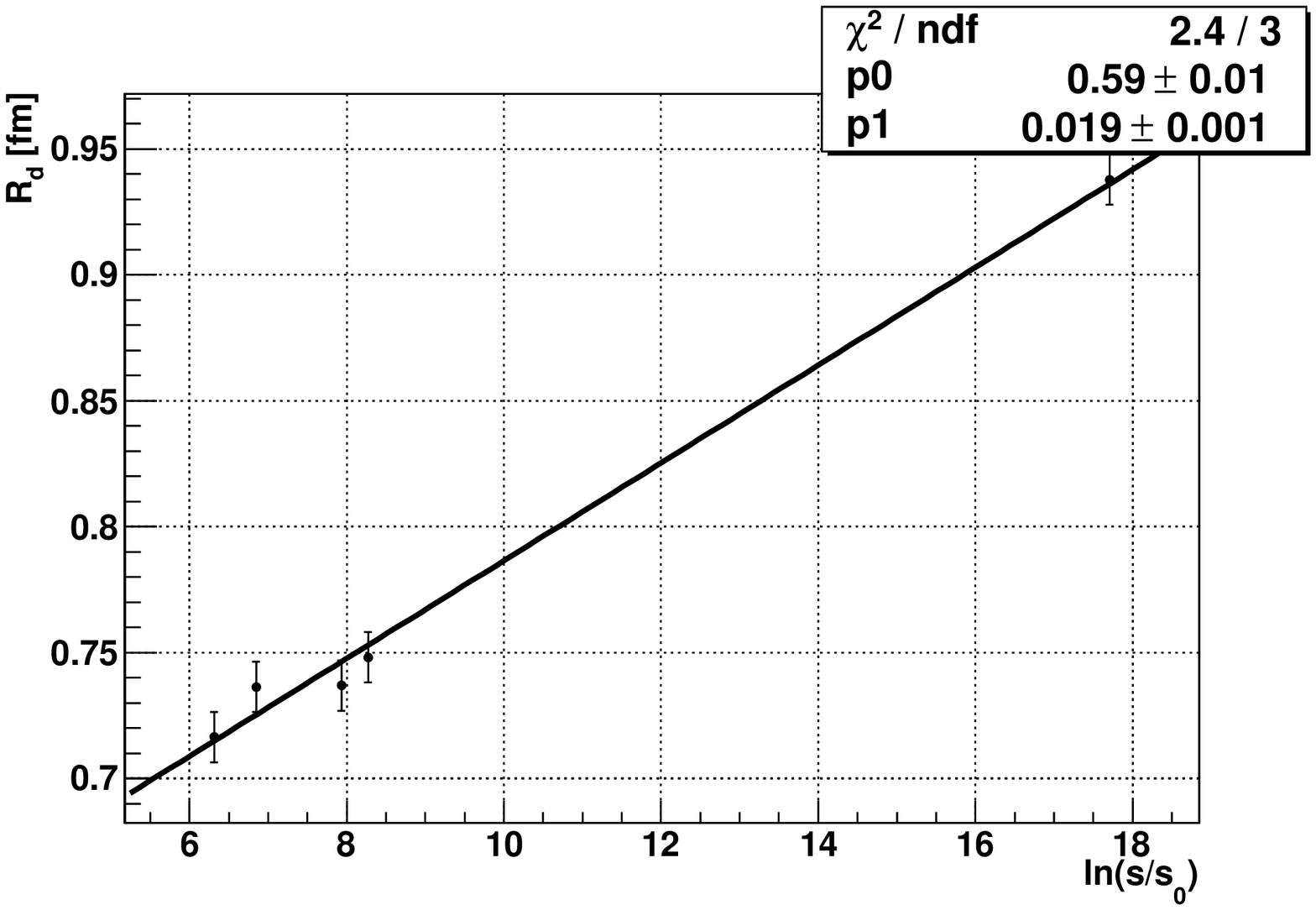}\\
\includegraphics[width=0.495\linewidth]{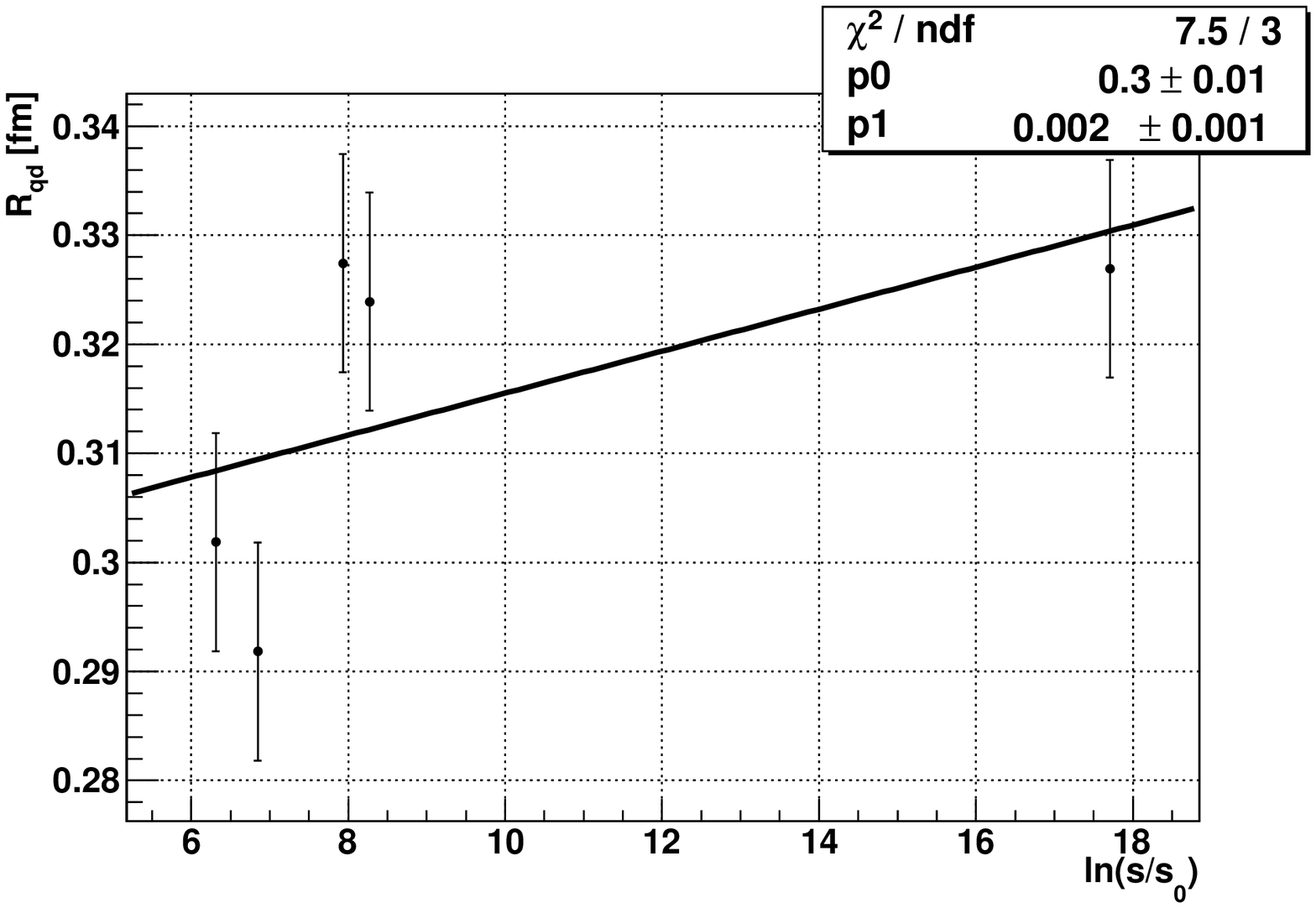}
\includegraphics[width=0.495\linewidth]{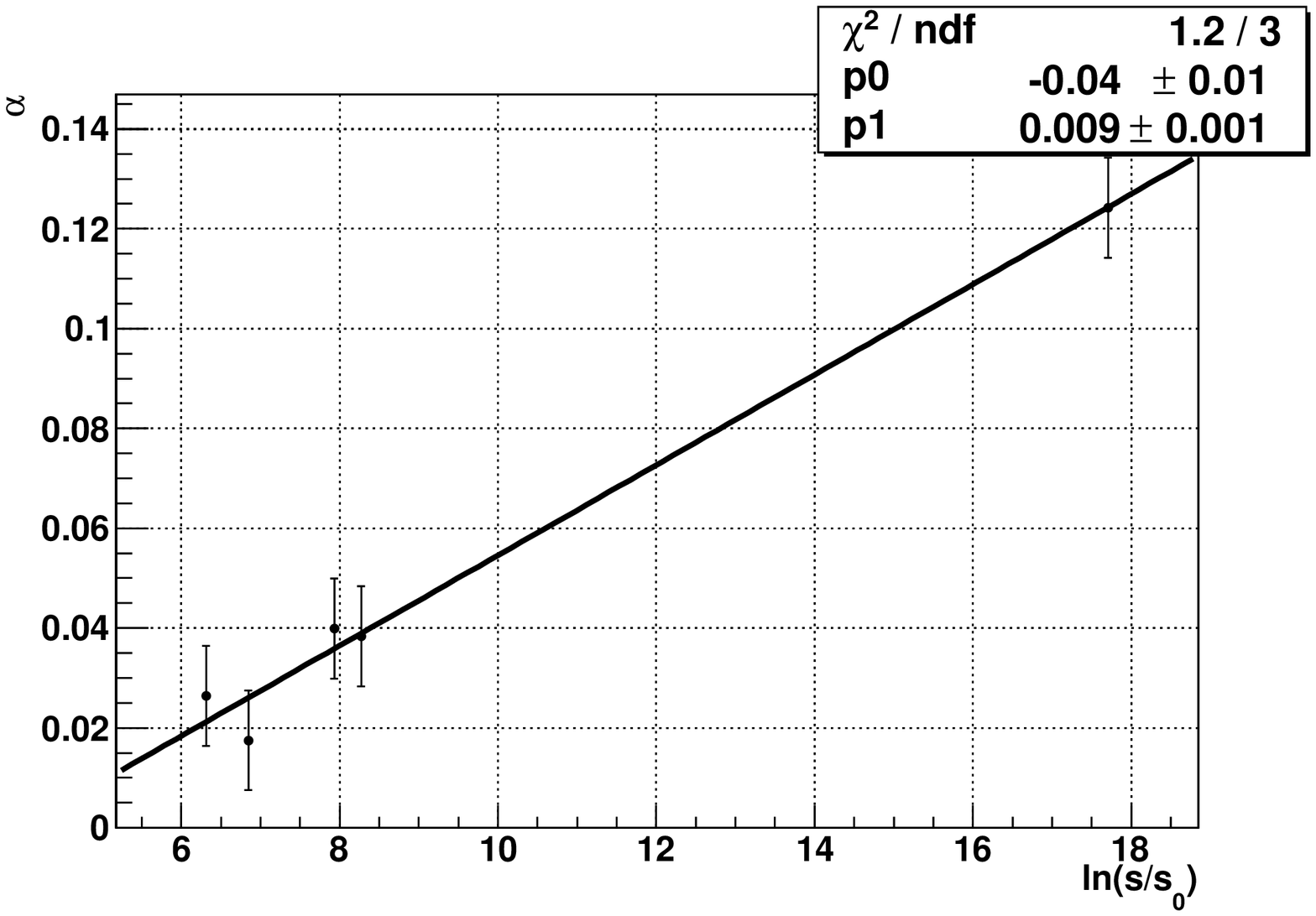}
\caption{
The excitation function of the parameters
of the ReBB model, collected in Table \ref{table:fit_parameters}, is determined
from fits  with Eq.~(\ref{parametrization_of_extrapolation}) to each of the
parameters $R_{q}$, $R_{d}$, $R_{qd}$ and $\alpha$. The plots about the
resulting fits are collected here, while the parameters of the
excitation functions are collected in Table~\ref{extrapolation_parameters}.
The statistically acceptable quality of these fits allow the ReBB model to be
extrapolated to center-of-mass energies which have not been measured yet at
LHC.}
\label{BBm_model_extrapolation_fits}
\end{figure}

	The ReBB model can be extrapolated to energies which have not been
measured yet at LHC. The fit results of Table \ref{table:fit_parameters} and
the parametrization  \begin{equation} P(s)=p_{0} + p_{1}\cdot\ln{(s/s_{0})}
\label{parametrization_of_extrapolation} \end{equation} is applied for each
parameter $P\in{\{R_{q},R_{d},R_{qd},\alpha\}}$, where $s_{0}=1$ GeV$^{2}$. The
parametrization Eq.~(\ref{parametrization_of_extrapolation}) implies that the
four free parameters of the original ReBB model that were fitted
at {\it each} colliding energy independently, corresponding to altogether 20
free fit parameters at the 5 energies analyzed in this manuscript are now
replaced with eight parameters $p_{i}$ that prescribe their
energy dependence. These fits to the energy dependence of the 
ReBB model parameters are shown in Fig.~\ref{BBm_model_extrapolation_fits} and
the fit parameters are collected in Table~\ref{extrapolation_parameters}.

The logarithmic dependence of the geometric parameters on the
center-of-mass energy $\sqrt{s}$ in the parametrization
Eq.~(\ref{parametrization_of_extrapolation}) is motivated by the so-called 
``geometric picture`` based on a series of studies~\cite{Cheng:1969eh,Cheng:1969bf,Cheng:1969ac,Chengbook,Bourrely:1978da,Bourrely:2014efa}.
In case of the $\alpha$ parameter, which is not a geometrical
property of the proton, the logarithmic $\sqrt{s}$ dependence is an additional
assumption. 
As indicated by Fig.~\ref{BBm_model_extrapolation_fits} and
Table~\ref{extrapolation_parameters}, 
such an energy dependence of the $\alpha$ shape
parameter is consistent with the currently available data.
 
Table~\ref{extrapolation_parameters} shows that the rate of increase with
$\sqrt{s}$, parameter $p_{1}$, is an order of magnitude larger for $R_{q}$ and
$R_{d}$ than for $R_{qd}$. The saturation effect, described in
Section~\ref{sec:shadow_profile}, is consistent with this observation as the
increasing components of the proton, the quark and the diquark, are confined
into a volume which is increasing more slowly.

\begin{table}[H]
\centering

\begin{tabular}{|c|c|c|c|c|} \hline
Parameter & $R_{q}$ [$fm$] & $R_{d}$ [$fm$] & $R_{qd}$ [$fm$]	& $\alpha$ \\ \hline\hline
$\chi^{2}/NDF$ 	& $6.2/3$& $2.4/3$ & $7.5/3$ 	& $1.2/3$ \\	\hline
CL [\%]		& 10.2	& 49.4     & 5.8 	& 75.3 \\	\hline
$p_{0}$ & $0.15\pm0.01$	  & $0.59\pm0.01$ & $0.30\pm0.01$ & $-0.04\pm0.01$ \\ \hline 
$p_{1}$ & $0.017\pm0.001$ & $0.019\pm0.001$& $0.002\pm0.001$& $\phantom{-}0.009\pm0.001$ 	\\   \hline
\end{tabular}

\caption{
The fits of the ReBB model to data at each ISR energy and at the 7~TeV LHC energy, summarized in Table \ref{table:fit_parameters}, allow for an extrapolation of the model parameters as a function of the center-of-mass
energy $\sqrt{s}$. 
The parametrization Eq.~(\ref{parametrization_of_extrapolation}) is applied to 
each parameter of the ReBB model
and the fits are shown in Fig.~\ref{BBm_model_extrapolation_fits}. 
Numerical values are rounded up to two valuable decimal digits.
The fit quality information is provided in the first and second row
of the table. Note that the fit quality is acceptable for each parameter
as CL$>$0.1 \% for each fit.}
\label{extrapolation_parameters}
\end{table}

%\vspace{-5mm}
\begin{wrapfigure}{r}{0.5\textwidth}
\vspace{-26pt}
\begin{center}
\includegraphics[trim = 0mm 7mm 0mm 20mm, clip, width=0.5\textwidth]{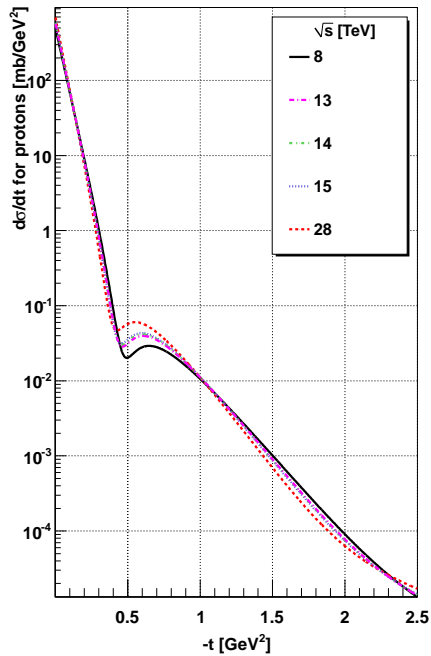}
\end{center}
\caption{The $pp$ elastic differential cross-section is extrapolated to 8 TeV as well as to future LHC energies and beyond.}
\label{BBm_model_extrapolation}
%\vspace{-30pt}
\end{wrapfigure}	

\begin{figure}[H] \centering
\includegraphics[width=0.495\linewidth]{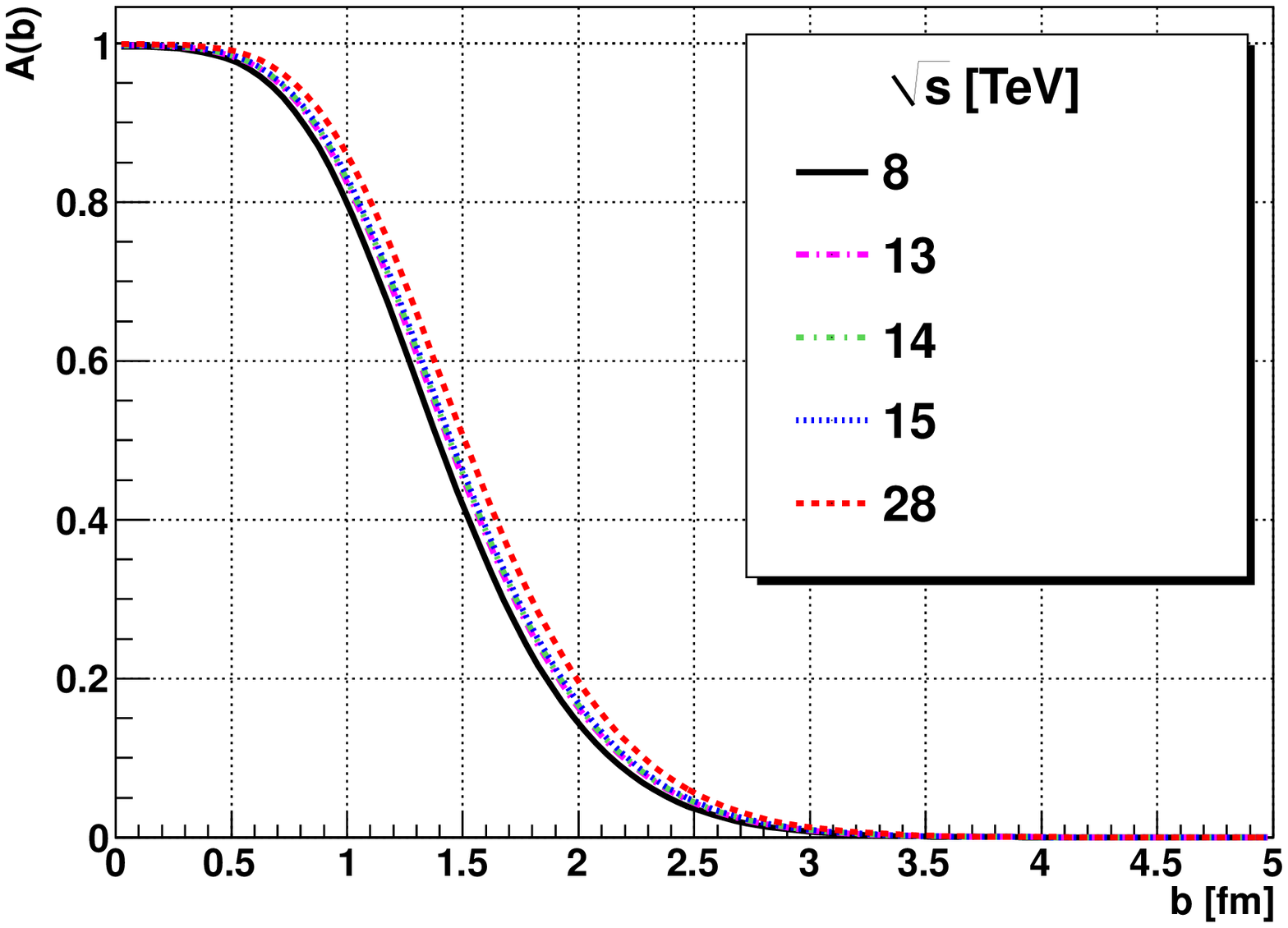}
\includegraphics[width=0.495\linewidth]{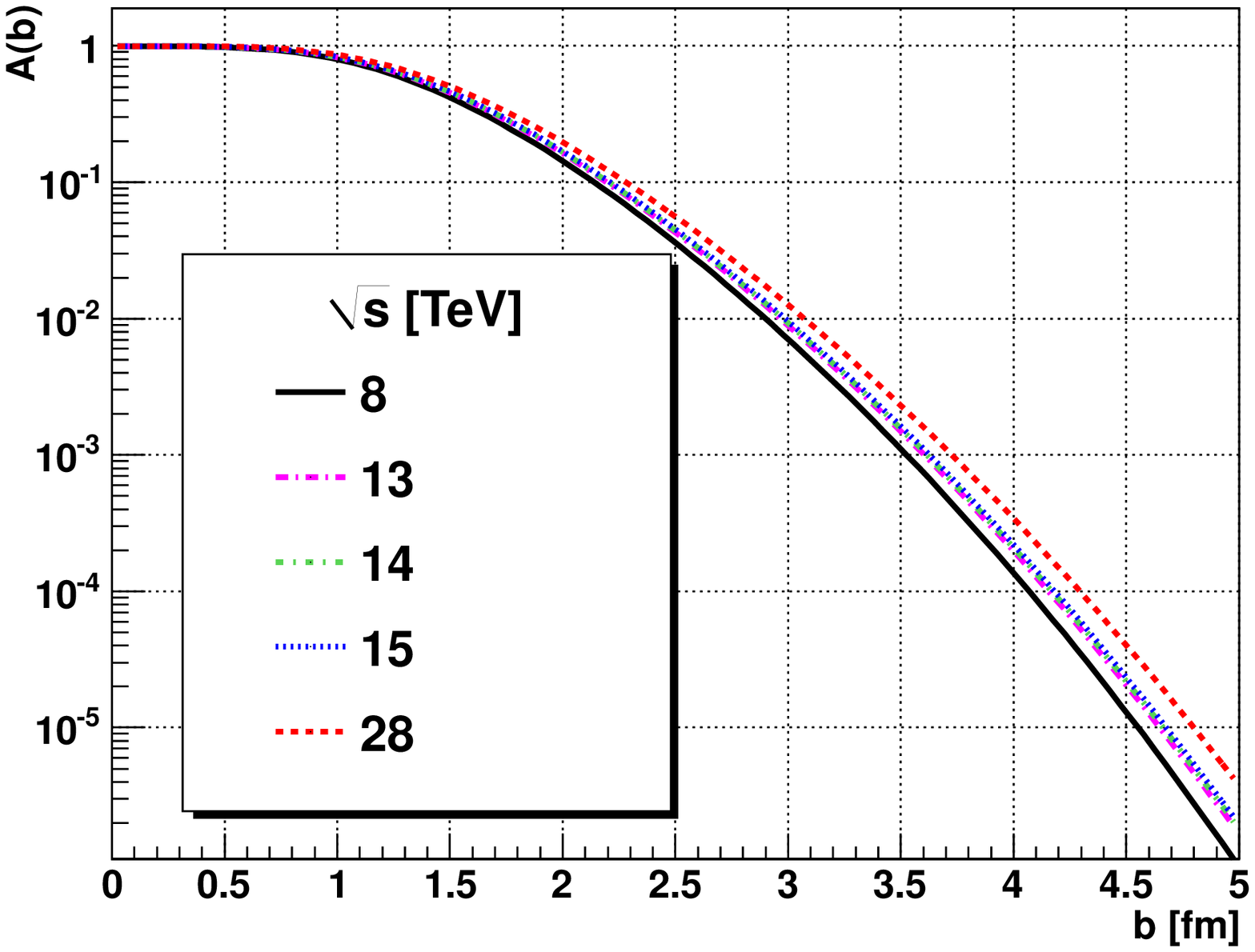}
\caption{The shadow profile function at the extrapolated energies $\sqrt{s}$.
The results show the increase of the proton interaction radius with increasing
$\sqrt{s}$ energies. Also note that the ``edge'' of the distributions remains
of approximately constant width and shape.}
\label{BBm_model_extrapolation_shadows} \end{figure}

\begin{figure}[h] \centering
\includegraphics[width=0.49\linewidth]{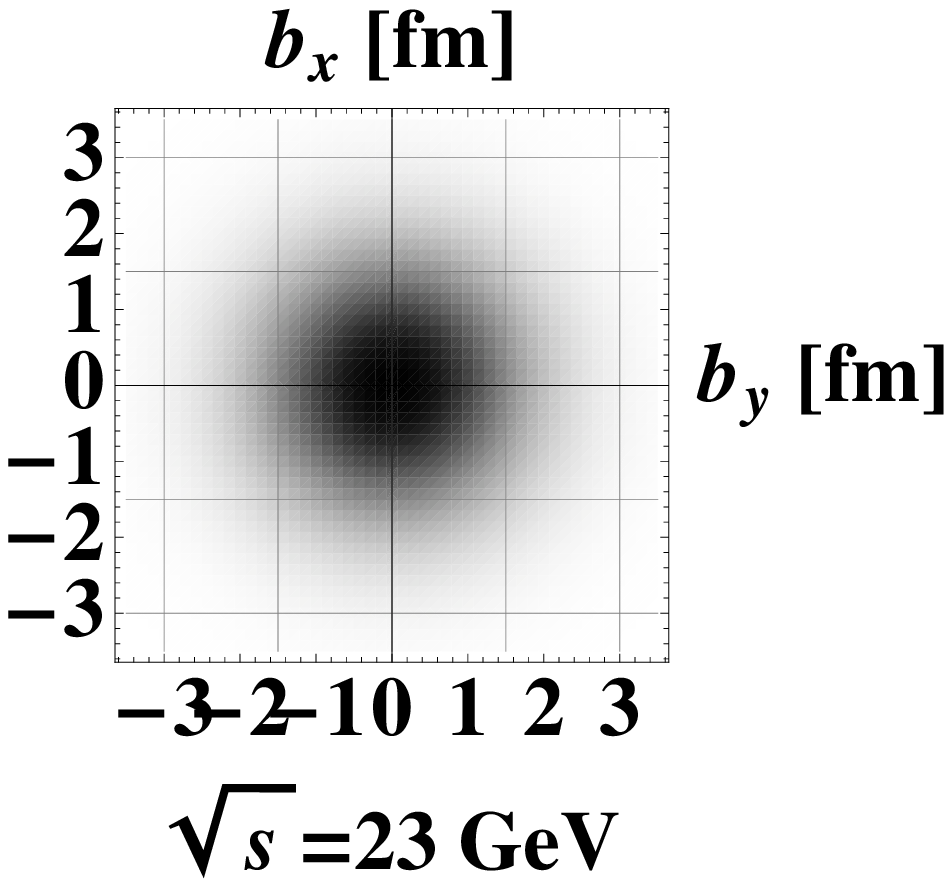}
\includegraphics[width=0.49\linewidth]{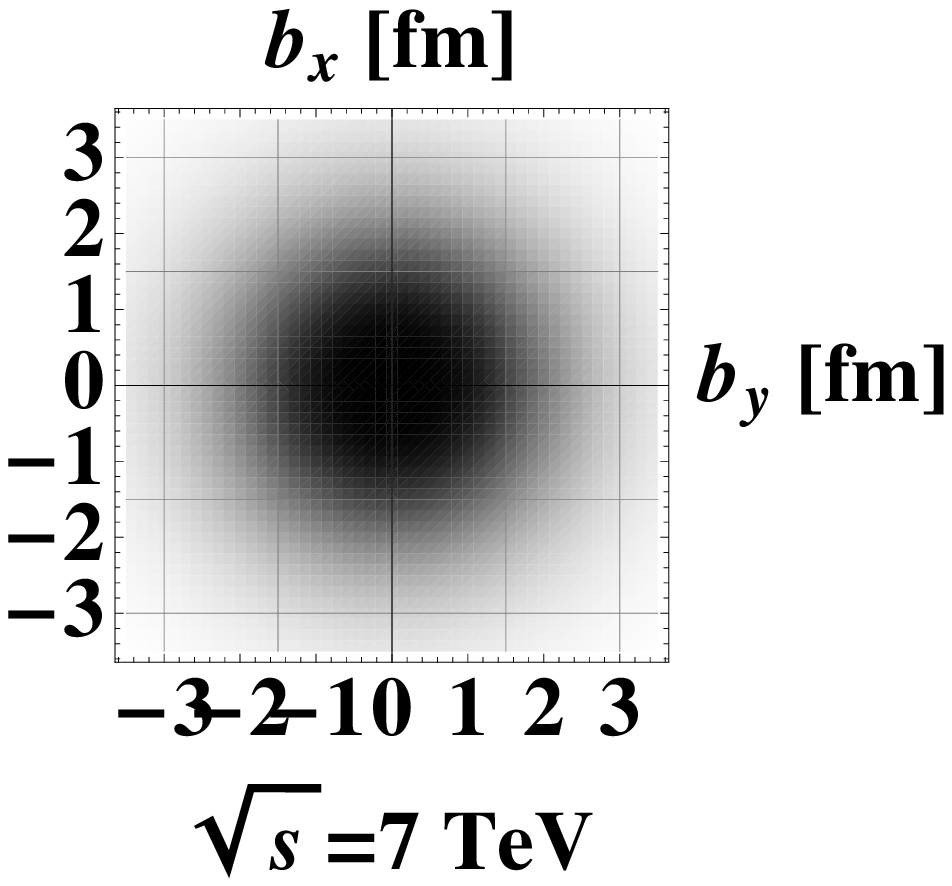}
\includegraphics[width=0.49\linewidth]{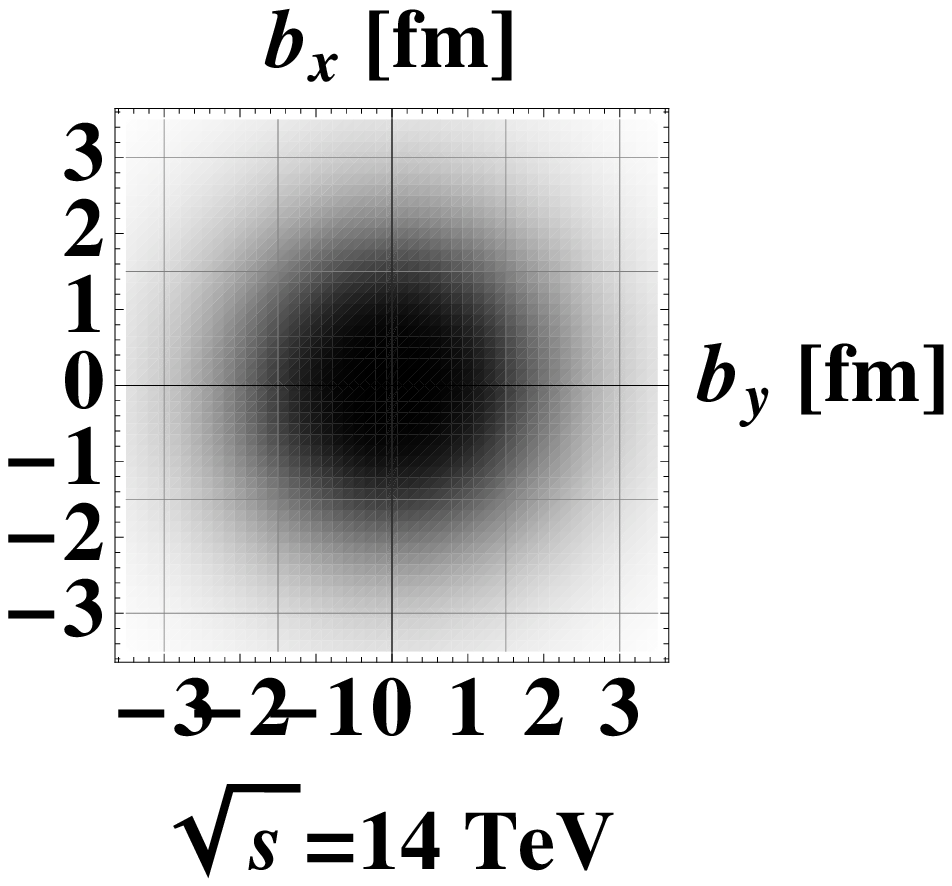}
\includegraphics[width=0.49\linewidth]{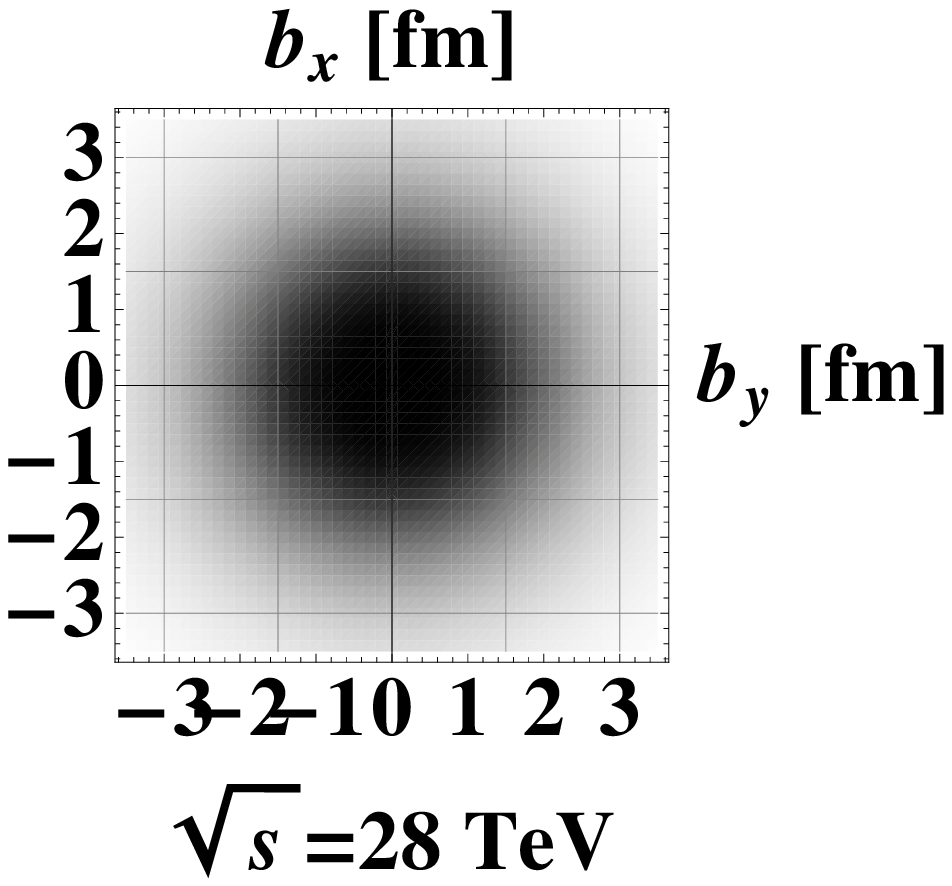} 
\caption{Visualization of the shadow profile functions $A(b)$ in the transverse
plane of the impact parameter vector~$(b_{x},b_{y})$. The figures show the
increase of the proton effective interaction radius in the impact parameter
space with increasing center-of-mass energy~$\sqrt{s}$. It can be also observed
that the black innermost core of the distributions is increasing, while the
thickness of the proton's ``skin'', the gray transition part of the
distributions, remains approximately independent of the center-of-mass energy
$\sqrt{s}$.} 
\label{BBm_model_fit_results_shadow_visualization}
\end{figure}

Using the extrapolation formula
Eq.~(\ref{parametrization_of_extrapolation}) and the value of the parameters
from Table~\ref{extrapolation_parameters} it is straightforward to calculate
the values of the parameters at 8 TeV, where the TOTEM
measurement of the total cross-section is published~\cite{Antchev:2013paa},
and at expected future LHC energies of $\sqrt{s}=$13, 14, 15 TeV and also at 28
TeV, which is beyond the LHC capabilities. Using the extrapolated values of the
parameters we plot our predicted $pp$ elastic differential cross-section curves
at each mentioned energy in Fig.~\ref{BBm_model_extrapolation}.  The shadow
profile functions $A(b)$ can be also extrapolated, see
Fig.~\ref{BBm_model_extrapolation_shadows}.  The shadow profile functions even
allow us to visualize the increasing effective interaction radius of the proton
in the impact parameter space in 
Fig.~\ref{BBm_model_fit_results_shadow_visualization}. 

	It is also important to see how the most important features change with
center-of-mass energy~$\sqrt{s}$: the extrapolated values of the total
cross-section $\sigma_{tot}$, the position of the first diffractive minimum
$|t_{dip}|$ and the parameter $\rho$ is given in Table
\ref{extrapolated_values}.

Our calculated value at $\sqrt{s}=8$~TeV is $\sigma_{tot}=99.6$~mb, which is consistent with the total cross-section $\sigma_{tot}=101.7\pm2.9$~mb at
$\sqrt{s}=8$~TeV, measured with a luminosity-independent method by the TOTEM experiment~\cite{Antchev:2013paa}. 
  
According to Table~\ref{extrapolated_values}, the predicted value of
$|t_{dip}|$ and $\sigma_{tot}$ moves more than 10\% when
$\sqrt{s}$ increases from 8~TeV to 28~TeV, while the value of
$C=|t_{dip}|\cdot\sigma_{tot}$ changes only about 2~\%, which
is an approximately constant value, within the errors of the extrapolation.

\begin{table}[H]
\centering
\begin{tabular}{|c||c|c|c|c|} \hline
$\sqrt{s}$ [TeV] & $\sigma_{tot}$ [mb] & $|t_{dip}|$[GeV$^{2}$] & $\rho$ & $|t_{dip}|\cdot\sigma_{tot}$ [mb~GeV$^{2}$] \\   \hline\hline
8        & 99.6         & 0.494 & 0.103 & 49.20 \\   \hline
13       & 106.4        & 0.465 & 0.108 & 49.48 \\   \hline
14       & 107.5        & 0.461 & 0.108 & 49.56 \\   \hline
15       & 108.5        & 0.457 & 0.109 & 49.58 \\   \hline
28       & 117.7        & 0.426 & 0.114 & 50.14 \\   \hline
\end{tabular}

\caption{The extrapolated values of the total cross-section $\sigma_{tot}$ at
future LHC energies and beyond. The position of the first diffractive minimum
$|t_{dip}|$, the parameter $\rho$ and the $|t_{dip}|\cdot\sigma_{tot}$ value is
also provided at each energy. Note that the predicted value of $|t_{dip}|$ and
$\sigma_{tot}$ moves more than 10\% when $\sqrt{s}$ increases from 8~TeV to
28~TeV, while the value of $|t_{dip}|\cdot\sigma_{tot}$ changes only about 2~\%.} 
\label{extrapolated_values} 
\end{table}

A similar, and exact, scaling can be derived for the case of photon scattering on a black disk, where the elastic differential cross-section is~\cite{Block:2006hy}
\begin{align}
\frac{d\sigma_{black}}{dt}=\pi R^{4}\left[\frac{J_{1}(q\cdot R)}{q\cdot R}\right]^{2}\,,
\label{dssigmadt_black}
\end{align}
where $R$ is the radius of the black disk and $t=-q^{2}$. The total cross-section is given by
\begin{align}
	\sigma_{tot,black}=2\pi R^{2}\,.  \label{sigmatot_black}
\end{align}
	
In this simple theoretical model the position of the first diffractive minimum, following from Eq.~(\ref{dssigmadt_black}), and the total cross-section Eq.~(\ref{sigmatot_black}) satisfies
\begin{align} C_{black}=|t_{dip,black}|\cdot\sigma_{tot,black}=2\pi
j_{1,1}^{2}(\hbar c)^2\approx~35.9\,\text{mb GeV}^{2}\,, \label{Cblack}
\end{align} 
where $j_{1,1}$ is the first root of the first order
Bessel-function of the first kind~$J_{1}(x)$.

The scaling behavior, indicated by the stability of the value $C$, 
is observed, but it is significantly 
different from the black disk model, described by Eq.~(\ref{Cblack}), 
as the corresponding value $C_{black}$ is significantly different
\begin{align}
		C_{black}\ne C\,.  \label{CblackCexp}
\end{align}
In this sense the value of $C$ indicates a more complex scattering phenomena, 
than the scattering of a photon on a black disc,
however, the constancy of the product suggests the validity of an asymptotic
geometric picture, in agreement with the recent observations in 
Refs.~\cite{Bautista:2012mq,CsorgO:2013kua,Giordano:2013iga,Anisovich:2014wha}.

\section{Summary and conclusions}

The real part of the forward scattering amplitude (FSA) is derived from
unitarity constraints in the Bialas-Bzdak model leading to the so-called ReBB
model. The added real part of the FSA significantly improves the model ability
to describe the data at the first diffractive minimum. In total the ReBB model
describes both the ISR and LHC data in the $0<|t|<2.5$~GeV$^{2}$ squared
four-momentum transfer range in a statistically acceptable manner; in the latter
case the fit range has to be divided to two parts, according to the compilation
of the two independent TOTEM measurements. The results are collected in
Table~\ref{table:fit_parameters}.

The fit results also permit us to evaluate the shadow profile functions $A(b)$,
see Fig.~\ref{BBm_model_fit_results_shadow}. The plots indicate a Gaussian
shape at ISR energies, while at LHC a saturation effect can be observed: the
innermost part of the shadow profile function $A(b)$ around $b=0$ is almost
flat and close to $A(b)\approx1$.  The elastic differential cross-section can
be compared to a purely exponential distribution and the comparison shows a
significant deviation from pure exponential in the $0.0\le|t|\le0.2$~GeV$^{2}$
range.

The fit results allow the determination of the excitation functions of the ReBB
model at future LHC energies and beyond, with parameters collected in Table 2
and predicted differential cross-section curves shown in
Fig.~\ref{BBm_model_extrapolation}. The shadow profile functions can be also
extrapolated, see Fig.~\ref{BBm_model_extrapolation_shadows}, which predicts
that the saturated part of the proton is expected to increase with increasing
center-of-mass energy $\sqrt{s}$. The edge of the distribution, the
``skin-width'' of the proton, expected to remain approximately constant. It is
worth to mention that the extrapolated version of the ReBB model utilizes of
only eight parameters, the $p_{i}$ parameters of
Table~\ref{extrapolation_parameters}, and in this sense a ``minimal`` set of
parameters is applied.

\section*{Acknowledgement} 
T. Cs. would like to thank to R. J. Glauber for inspiring and useful discussions 
and for kind hospitality during his visits to Harvard University. The authors are also
grateful to G.~Gustafson, L.~Jenkovszky, H.~Niewiadomski and J.~Ka\v{s}par
for inspiring and fruitful discussions. This work was partially supported by the OTKA
grant NK 101438 (Hungary) and the Ch.~Simonyi Fund (Hungary).

\end{document}